\documentclass[12pt,preprint]{aastex}
\usepackage{color}
\usepackage{graphics,graphicx}
\usepackage[caption=false,font=footnotesize]{subfig}
\usepackage{textcomp}
\usepackage[varg]{txfonts}
\usepackage{natbib}
\usepackage{arydshln}

\shorttitle{SEP events modeled on the Martian surface}
\shortauthors{Guo et al.}

\begin{document}

\title{A generalized approach to model the spectra and radiation dose rate of \\
solar particle events on the surface of Mars}

\author{
 Jingnan Guo\altaffilmark{1},
 Cary Zeitlin \altaffilmark{2},
 Robert~F. Wimmer-Schweingruber\altaffilmark{1},
 Thoren McDole\altaffilmark{1},
 Patrick K\"uhl\altaffilmark{1},
 Jan C. Appel\altaffilmark{1},
 Daniel Matthi\"a \altaffilmark{3},
 Johannes Krauss \altaffilmark{1},
 Jan K\"ohler \altaffilmark{1}
}

\altaffiltext{1}{Institute of Experimental and Applied Physics, Christian-Albrechts-University, Kiel, Germany\email{guo@physik.uni-kiel.de}}\label{inst:kiel}
\altaffiltext{2}{Leidos, Houston, Texas, USA} \label{inst:leidos}
\altaffiltext{3}{German Aerospace Agency, Cologne, Germany}

\begin{abstract}
	For future human missions to Mars, it is important to study the surface radiation environment during extreme and elevated conditions. In the long term, it is mainly Galactic Cosmic Rays (GCRs) modulated by solar activity that contributes to the radiation on the surface of Mars, but intense solar energetic particle (SEP) events may induce acute health effects. Such events may enhance the radiation level significantly and should be detected as immediately as possible to prevent severe damage to humans and equipment. However, the energetic particle environment on the Martian surface is significantly different from that in deep space due to the influence of the Martian
	atmosphere. Depending on the intensity and shape of the original solar particle spectra as well as particle types, the surface spectra may induce entirely different radiation effects. In order to give immediate and accurate alerts while avoiding unnecessary ones, it is important to model and well understand the atmospheric effect on the incoming SEPs including both protons and helium ions. In this paper, we have developed a generalized approach to quickly model the surface
	response of any given incoming proton/helium ion spectra and have applied it to a set of historical large solar events thus providing insights into the possible variety of surface radiation environments that may be induced during SEP events. Based on the statistical study of more than 30 significant solar events, we have obtained an empirical model for estimating the surface dose rate directly from the intensities of a power-law SEP spectra.
\end{abstract}


\section{Introduction and Motivation}\label{sec_intro}
In order to plan future human missions to Mars, the assessment of the radiation environment on and near the surface of Mars is necessary and fundamental for the safety of astronauts. 
There are two types of primary particles reaching the top of the atmosphere of Mars: galactic cosmic rays (GCRs) and solar energetic particles (SEPs). 
GCRs, mainly composed of protons and helium ions,  are modulated by heliospheric magnetic fields which evolve dynamically as solar activity varies in time and space, with a well-known 11-year cycle \citep[e.g.,][]{parker1958interaction}.
SEP events, consisting mainly of protons, are sporadic and highly variable in terms of their intensities and energy spectra. They take place much more frequently during solar maximum periods and they may enhance the radiation level significantly, and therefore should be detected as quickly as possible to minimize risks to humans and equipment on the Martian surface.

However, SEP measurements at Mars are very scarce and within a limited energy range. The radiation assessment detector (RAD) onboard the Mars Science Laboratory (landed on Mars in Aug 2012) has measured only 6 moderate events in the course of 5 years during the declining phase of the past solar maximum \citep{hassler2014}. The SEP instrument onboard the Mars Atmosphere and Volatile EvolutionN \cite[MAVEN/SEP,][]{larson2015maven} spacecraft orbiting Mars since October 2014 only directly measures protons with energies $\leq 6$ MeV which do not contribute to the surface radiation enhancement as will be shown in this study. 
At near-Earth environment, SEPs are measured much more frequently by particle detectors on various spacecraft such as the Solar and Heliospheric Observatory (SOHO), the Advanced Composition Explorer (ACE), the Geostationary Operational Environmental Satellite (GOES) and so on.  
To derive the particle spectra at Mars location from these measurements is however very challenging. 
This is because the propagation of coronal mass ejections (CMEs) and the associated shocks (which are believed to be a major accelerator for such highly-energetic particles) through the heliosphere may result in totally different particle spectral intensities and shapes at Mars compared to Earth \citep{li2003energetic}. Besides, the observed SEP spectra and intensity also depend on different magnetic connections of the planets/spacecraft to the acceleration locations. 
The current paper will not address the above issues when considering the SEP induced radiation environment on the surface of Mars. Alternatively we focus on how the primary energy spectra are influenced and modified by the Martian atmosphere considering the presence of some SEP events at Mars which have been observed at near-Earth locations.

The energetic particle environment on the Martian surface is different from that in deep space due to the presence of the Martian atmosphere. 
{Much work has been done for calculating the surface exposure rates under different GCR and SEP scenarios. 
Several models combining particle transport codes with different GCR and/or SEP spectra have been developed and applied for estimating the radiation exposure on the surface of Mars \citep[e.g.,][]{keating2005model, deangelis2006modeling, mckenna2012characterization, ehresmann2011}. \citet{saganti2004radiation} have mapped the radiation exposure on the Mars surface from GCRs. 
\citet{simonsen1990radiation} and \citet{simonsen1992mars} calculated the surface dose exposures from both GCRs during solar minimum and maximum conditions as well as some significant SEP events.
\citet{townsend2011estimates} considered the transport of possible Carrington-type solar energetic particle events through the Martian atmosphere and also through various hemispherical configurations of aluminum shielding to estimate the resulting organ doses and
effective doses of such extreme events.
{{\citet{norman2014influence} have investigated the influence of dust loading on atmospheric ionizing radiation during solar quiet and SEP events.}} 
\citet{dartnell2007modelling} have also estimated the effect of surface radiation on the likelihood of survival of microbial life in the Martian soil.}
 
Depending on the intensity and shape of the original solar particle spectra as well as the distribution of particle types, different SEP events may induce entirely different radiation effects on the surface.
This is because primary particles passing through the Martian atmosphere may undergo inelastic interactions with the ambient atomic nuclei creating secondary particles (via spallation and fragmentation processes), which may also interact while propagating further and finally result in very complex spectra including both primaries and secondaries at the surface of Mars \citep[e.g.,][]{saganti2002, guo2015modeling}. 
Primary particles with small energies do not have sufficient range to reach the ground, but the exact energy cutoff is a strong function of elevation on Mars. Therefore, an intense SEP spectra {with moderate high-energy component} could be well within biological tolerance seen on the surface of Mars, particularly in low-lying places such as Gale Crater, Hellas Planitia, Valles Marineris, etc., where atmospheric shielding is substantially greater than the global average.
 
In order to give immediate and precise alerts while avoiding unnecessary ones, it is important to model and well understand the atmospheric effect on the incoming SEPs {and how this effect depends on the properties of the incoming SEP}. 
There are various particle transport codes such as HZETRN \citep{slaba2016solar, wilson2016}, PHITS \citep{sato2013} and GEANT4/PLANETOCOSMICS \citep{desorgher2006planetocosmics} which can be employed for studying the particle spectra and radiation through the Martian atmosphere. 
\citet{gronoff2015computation} have applied both PLANETOCOSMICS and HZETRN to calculate the GCR radiation environment on the surface of Mars and found highly consistent results from both simulations. 
In this paper we use the PLANETOCOSMICS transport code and develop a generalized approach to quickly model the surface response of any given incoming proton spectrum under different atmospheric depths. 
We have further applied the method to a set of significant solar events which took place in the last several decades, thus providing insights into the possible variety of surface particle spectra and induced radiation environment during SEP events {which are not only worst-case scenarios but also less extreme and frequent ones}. Moreover, we have obtained an empirical model for estimating the SEP-induced surface dose rate directly from the intensities of a power-law shaped SEP event.

\section{A generalized model: PLANETOMATRIX}\label{sec:planetomatrix}

\begin{figure}[htb!]
\centering
\begin{tabular}{cc}
\subfloat[input: proton, output:  downward proton] { \includegraphics[trim=42 20 80 35,clip, scale=0.5]{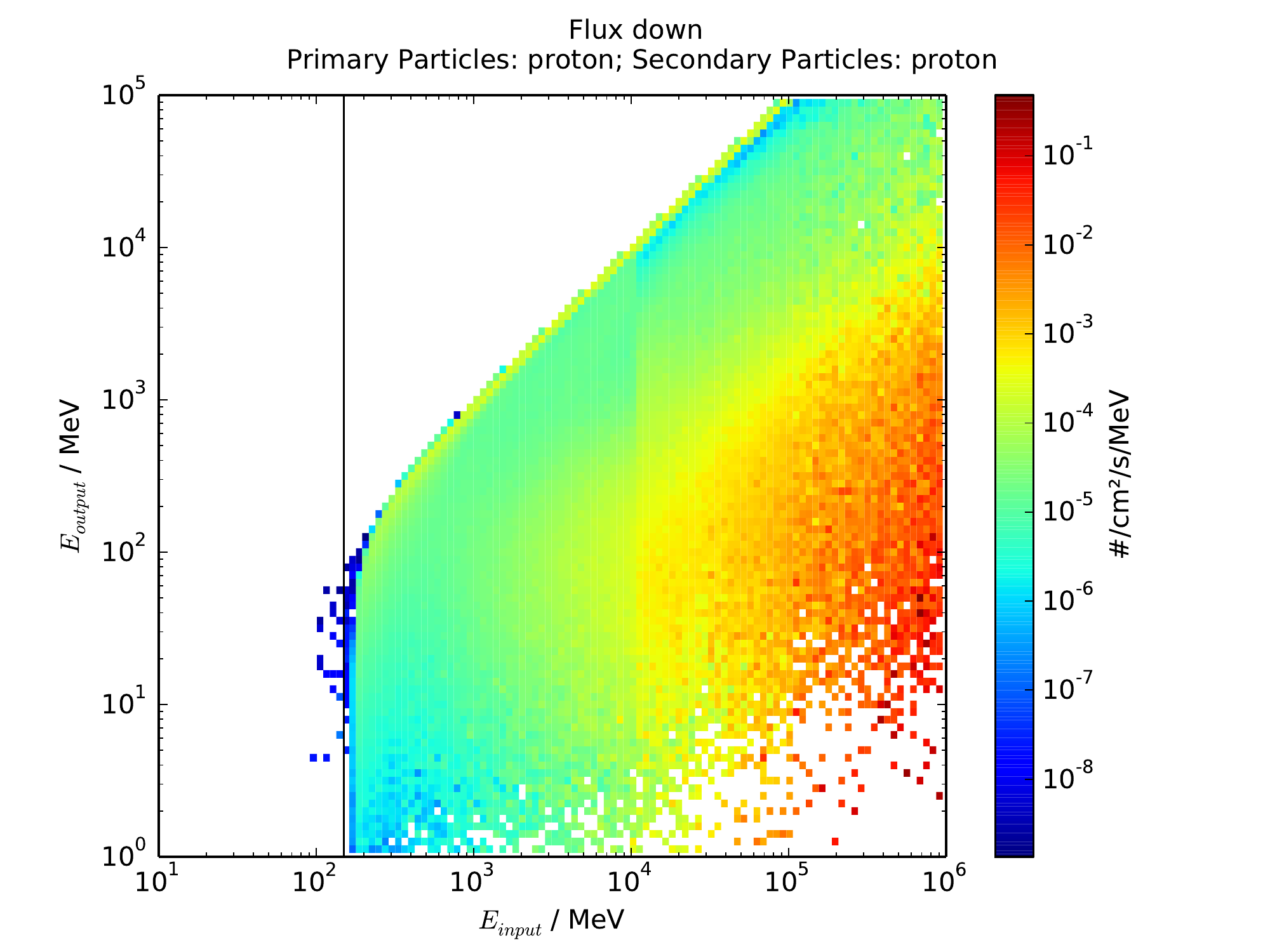} } & 
\subfloat[input: proton, output:  upward proton] { \includegraphics[trim=42 20 135 35,clip, scale=0.5]{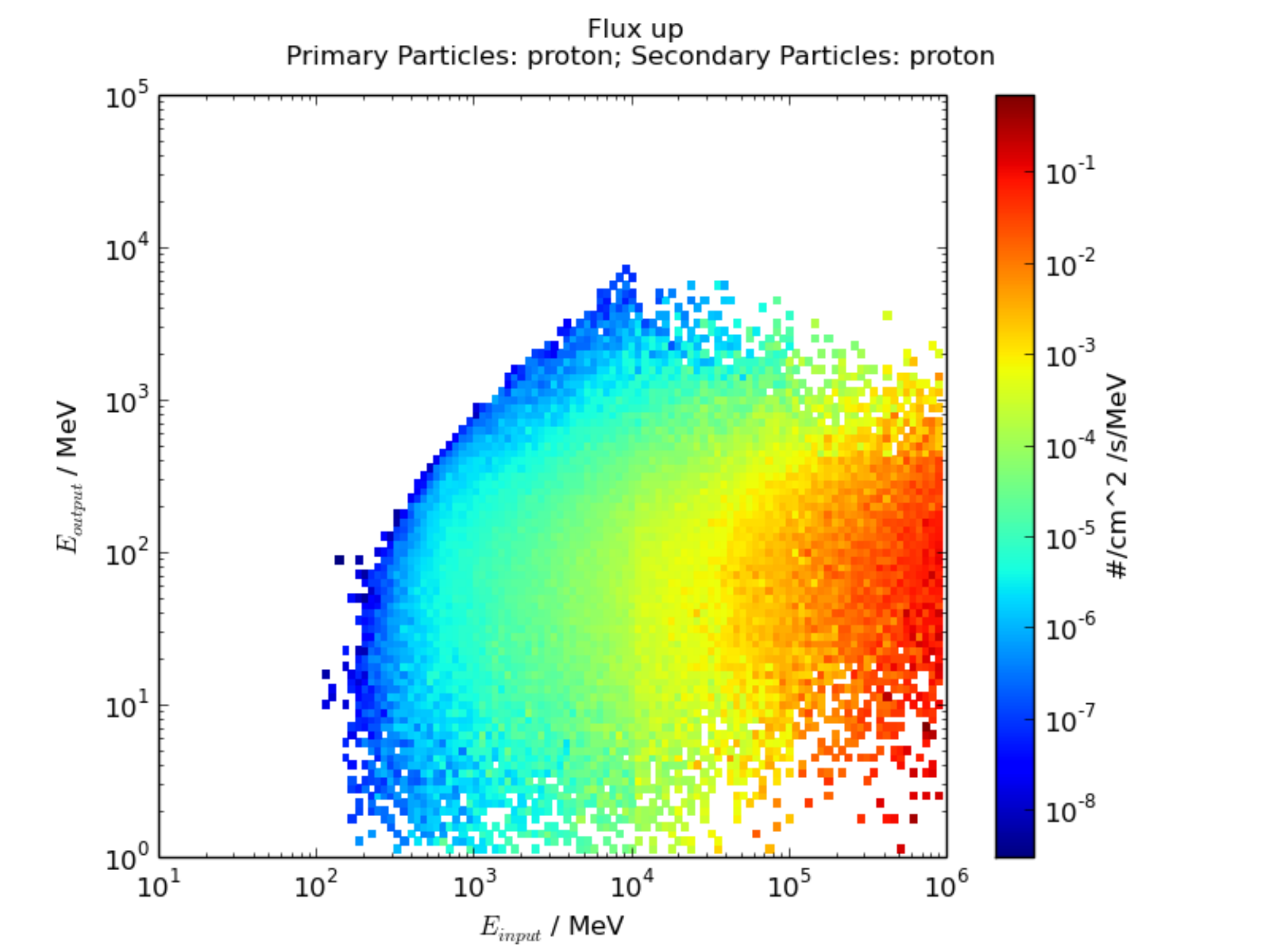}} \\
\subfloat[input: proton, output:  downward neutron] { \includegraphics[trim=42 20 135 35,clip, scale=0.5]{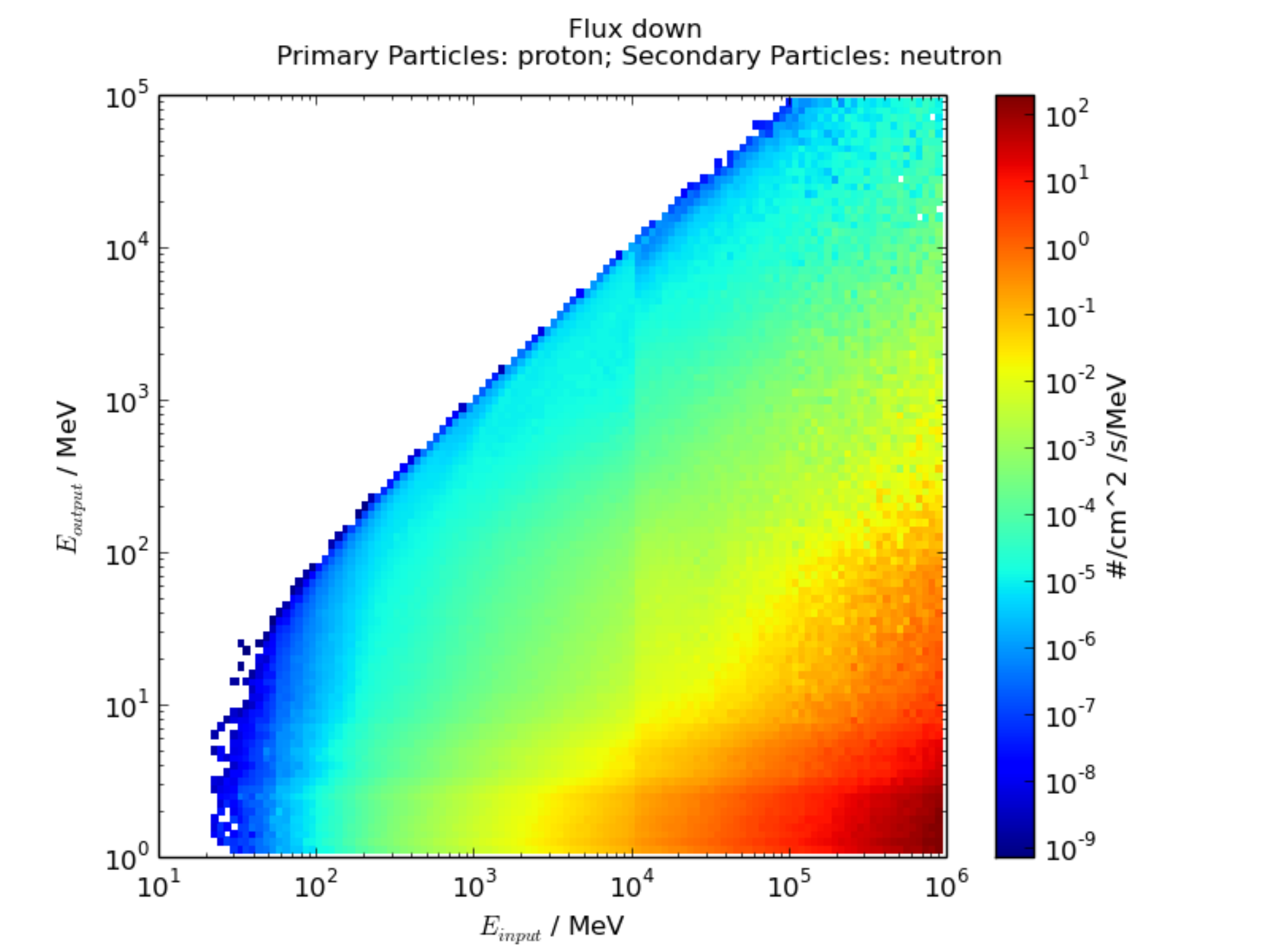} } & 
\subfloat[input: proton, output:  upward neutron] { \includegraphics[trim=42 20 135 35,clip, scale=0.5]{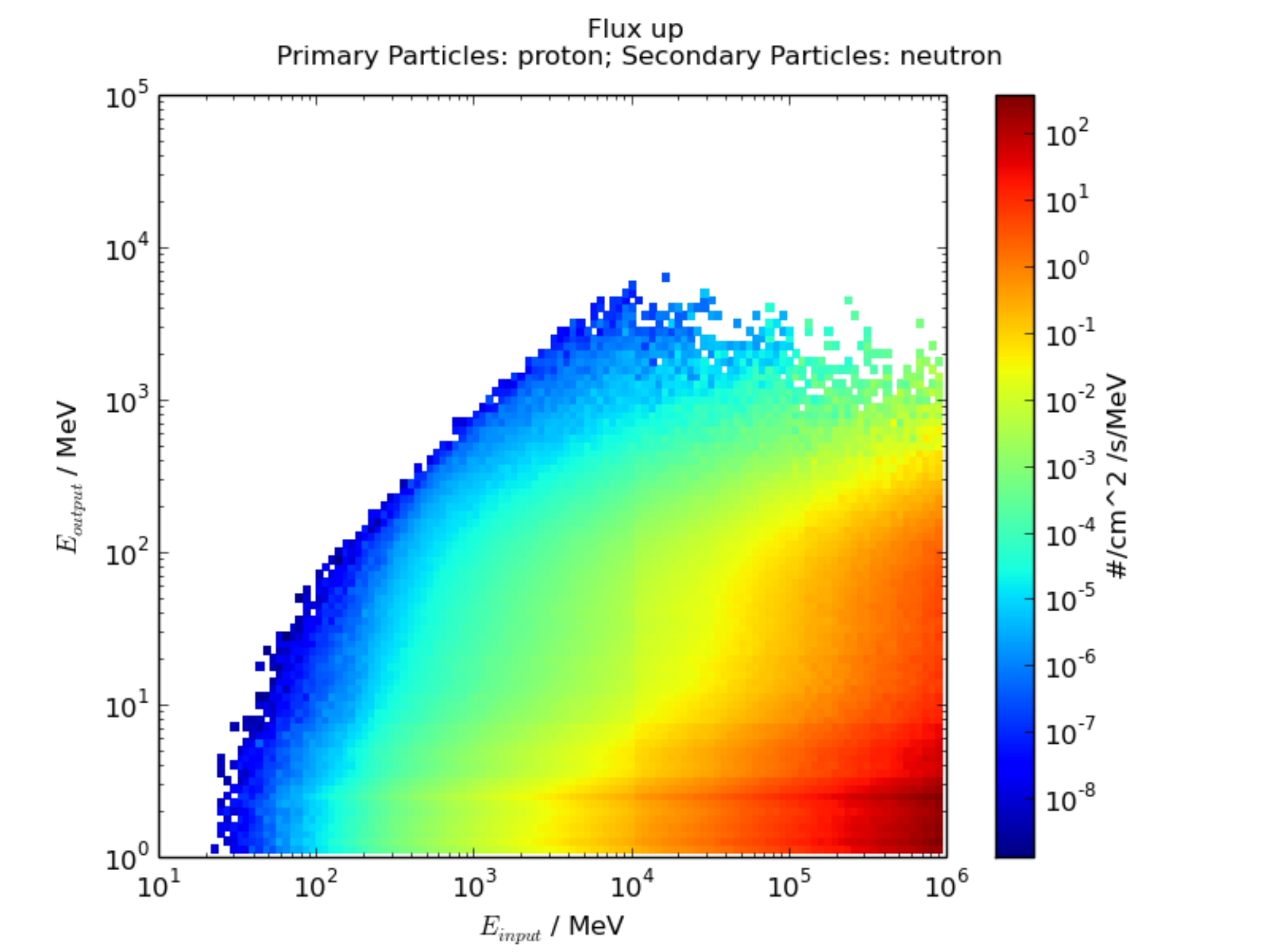}}
\end{tabular}
\caption{2-d histogram of (a) proton-downward proton matrix, (b) proton-upward proton matrix, (c) proton-downward neutron matrix, and (d) proton-upward neutron matrix under a vertical column depth of 20 g/cm$^2$. 
X-axis and Y-axis stand for the input and output energies [MeV] respectively. The vertical line in (a) marks the 150 MeV input proton energy.
The producing probabilities are represented by colors using a logarithmic color distribution. Each column in this plot is equivalent to a normalized output spectrum obtained by PLANETOCOSMICS using protons of the corresponding input energy. 
}\label{fig:PLANETOMATRIX}
\end{figure}

\begin{figure}[htb!]
\centering
\begin{tabular}{cc}
\subfloat[input: $^4$He, output:  downward $^4$He] { \includegraphics[trim=42 20 135 35,clip,scale=0.4]{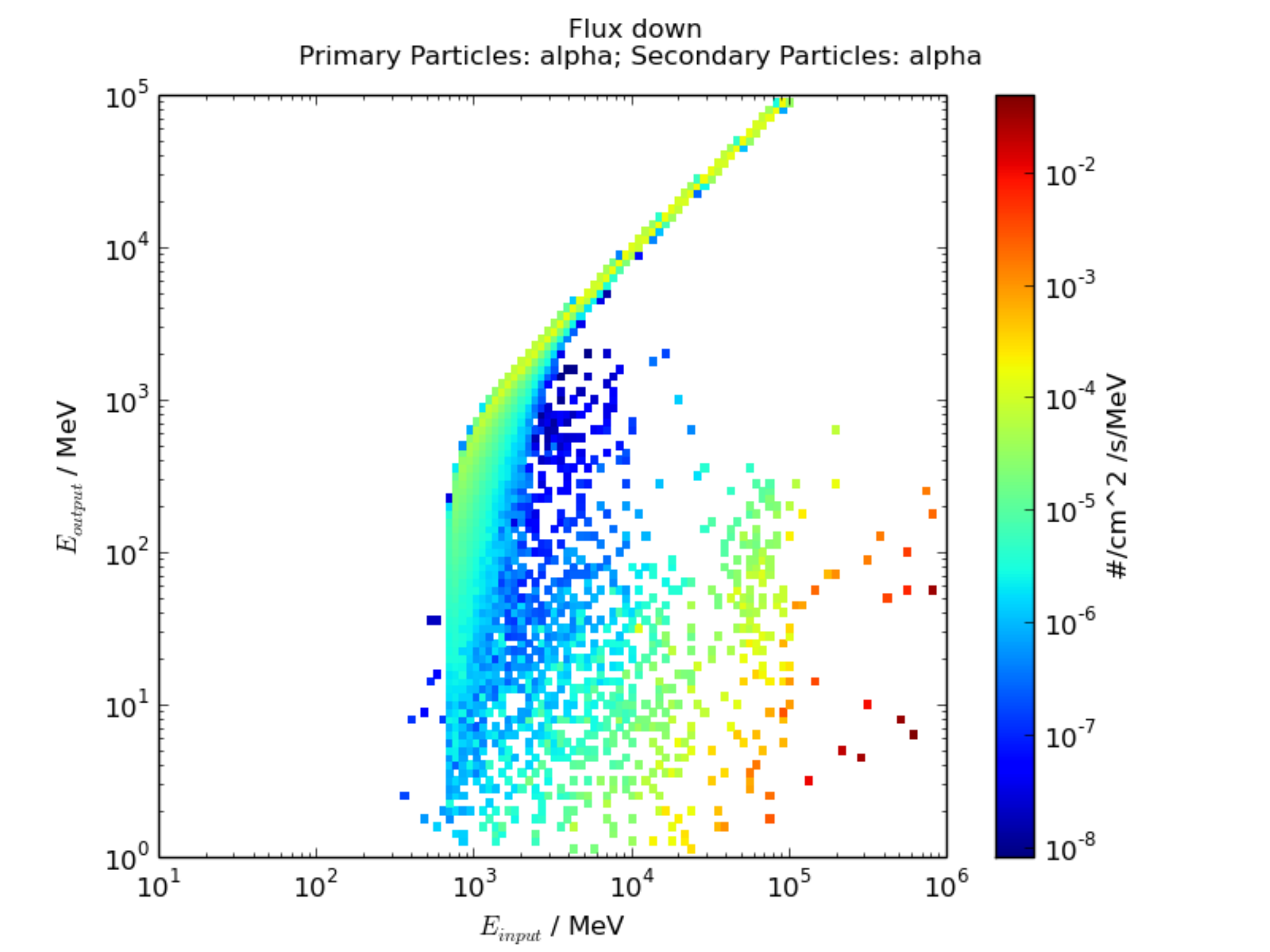} } & 
\subfloat[input: $^4$He, output:  downward proton] { \includegraphics[trim=42 20 80 35,clip,scale=0.4]{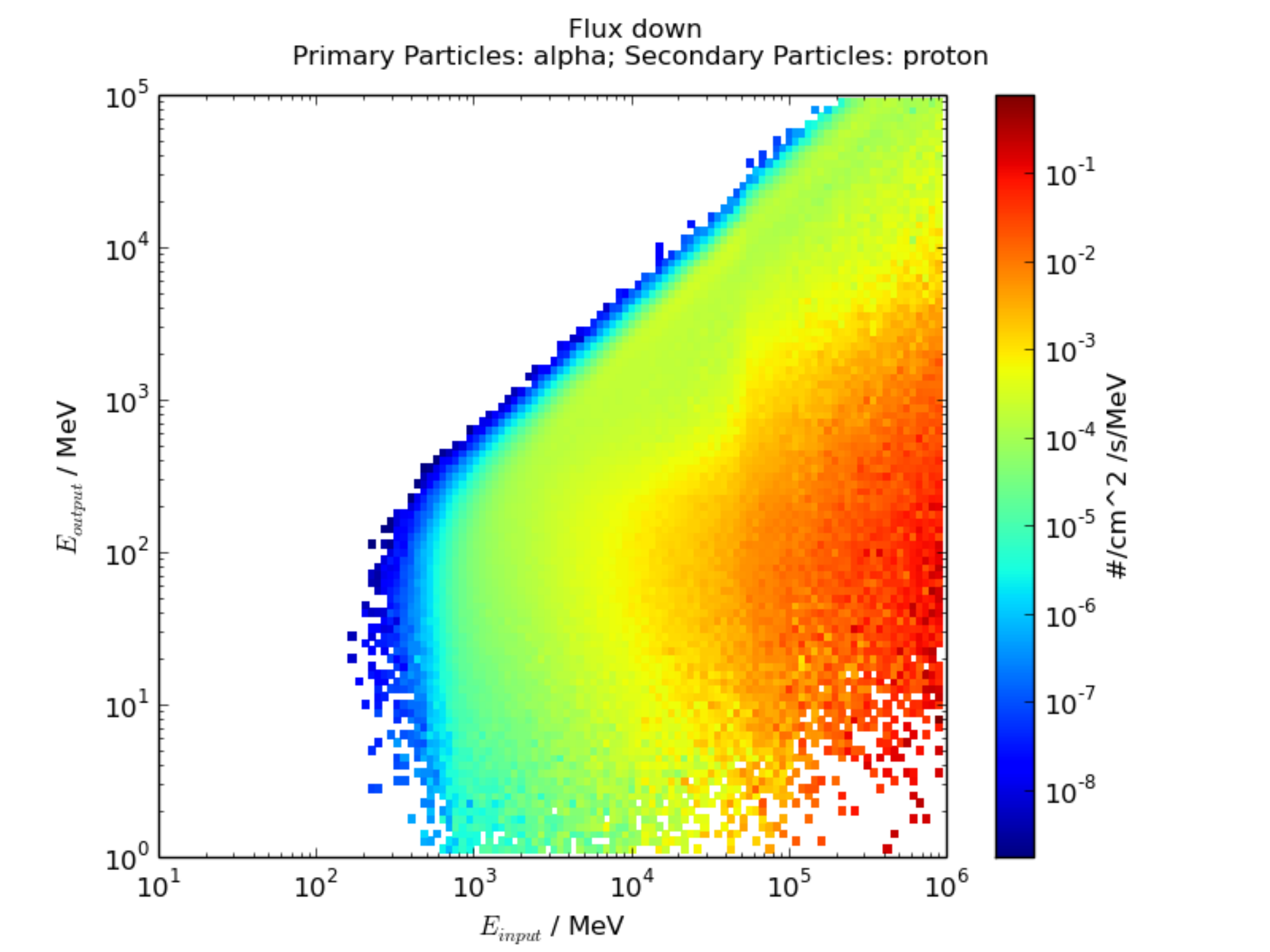}} 
\end{tabular}
\caption{2-d histogram of (a) $^4$He to downward $^4$He matrix and (b) $^4$He to fragmented downward proton matrix under a vertical column depth of 21 g/cm$^2$. 
X-axis and Y-axis stand for the input and output energies [MeV] respectively. 
The normalized intensities are represented by colors using a logarithmic color distribution.
}\label{fig:PLANETOMATRIX_alpha}
\end{figure}

PLANETOCOSMICS is a simulation tool (Version g4.10 has been used herein) developed in order to simulate particles going through planetary atmospheres and magnetic fields \citep{desorgher2006planetocosmics}.
It is based on GEANT4, a Monte Carlo approach for simulating the interactions of particles as they traverse matter \citep{agostinelli2003}. 
Different settings and features, e.g. the composition and depth of the atmosphere and the soil, can be used in the simulations.
Employing PLANETOCOSMICS to model of the radiation environment on the surface of Mars has been carried out in various studies \citep[e.g., ][]{dartnell2007modelling, gronoff2015computation, matthia2016martian, ehresmann2011} and has been validated when compared to spectra of the proton, helium ion and heavier ion spectra on the surface of Mars \citep{matthia2016martian} measured by the radiation assessment detector (RAD) onboard the Mars Science Laboratory (MSL).

In order to provide a more realistic atmospheric environment for the simulations, we use the Mars Climate Database (MCD) which has been created using different Martian atmospheric circulation models which are further compared and modified by the observation results from past and current Mars missions \citep{lewis1999climate}. 
It offers the possibility to access atmospheric properties, such as temperature, pressure and composition, for different altitudes, seasons and even the time of the day on Mars.
We use the composition, density and temperature profiles from MCD between altitudes of 250 and 0 km above the ground in steps of 100 m. A web interface of the MCD can be found here http://www-mars.lmd.jussieu.fr and used to assess the global map, daily and seasonal variation as well as the vertical profiles of the Martian atmosphere. 
In our simulations, we implement the MCD atmosphere properties at the location of Gale Crater (latitude: -4.6 and longitude 250) which is the landing site of the Curiosity rover \citep{grotzinger2012mars}. 

A full PLANETOCOSMICS simulation can be highly time-consuming and in principle needs to be run for each different input spectra. To reduce the computational burden, we developed an alternative approach which we refer to as the PLANETOMATRIX method, which folds the complicated nuclear interaction process into a two-dimensional matrix $\rm{\bar A(E_0, E)}$ where $E_0$ is the energy of a particle above the Martian atmosphere and $E$ is the particle energy on the Martian surface. 
It is constructed in the following way: First, a primary particle spectrum $f_m$ in the range of a single energy bin $E_{0m}$ (e.g. from 200 MeV to 210 MeV) is fed into the PLANETOCOSMICS code to generate the surface spectrum which is different from the original due to the production of secondaries. This surface spectrum can be described by a histogram with N bins and the flux in each bin $n$ is $a_{mn}$ (normalized to the input flux $f_m$). 
Second, this process is repeated M times for M different input energy bins (covering 1 to 10$^6$ MeV of primary particle energy) and the resulting scaled fluxes in each $(m,n)$ bin are $a_{mn}$. 
Thus, under a given atmospheric composition and column depth $\sigma$ setup, the matrix $\rm{\bar A(E_0, E)}$ (with a shape of $\rm{M} \times N$) can be constructed by running M simulations of PLANETOCOSMICS. 
Finally with an input spectrum $f(E_0)$ at the top of the atmosphere, the surface spectrum can be calculated as $\rm{F(E)} = \rm{\bar A(E_0, E)} \cdot f(E_0)$. 

{Different physics lists describing inelastic hadronic (nucleon or nuclear) interactions and electromagnetic interactions in PLANETOCOSIMICS have been tested extensively and compared with MSL/RAD measurement \citep[][Table 2]{matthia2016martian} . It was found that the selection of different physical model lists in GEANT4 in most cases (especially for low-Z particles) does not affect the resulting radiation exposure significantly with a maximum difference of 20\% in the dose equivalent rate. The simulation setup used here has employed the "emstandard opt4" model for electromagnetic interactions and the Binary intranuclear cascade model "QGSP BIC HP" to calculate cross section for $\sim$ GeV nucleons \citep{ivanchenko2004geant4}. 
{{The cut-in-range for electrons, positrons and gammas are set to be 0.5 g/cm$^2$ below which they are not tracked any longer. The minimum tracking energies for electrons and protons are 1 MeV and for neutrons and gammas are 1 keV and 100 keV respectively. A comparison with the results for lower energy cuts in \citet{gronoff2012computing} demonstrated that the resulting dose is hardly affected by this approximation.}}
Our PLANETOMATRIX folds in the process of primary interactions with the Martian atmosphere and regolith and generations of secondaries therein and describes how a given input spectrum is modified to produce output spectra of different secondaries through this process.} 
It is a statistical description, i.e., both f (with M bins) and F (with N bins) are energy-dependent distribution histograms and each element in the matrix $\rm{\bar A(E_0, E)}$ represents the probability of a primary particle with energy $E_0$ resulting in a particle on the surface with energy $E$.

Although the construction of each matrix is time-consuming, the multiplication of different input spectra with such a matrix to generate different surface spectra is very much simplified. 
Furthermore, with measurements of surface spectra $\rm{F(E)}$ by, e.g, MSL/RAD, an inversion technique can, in principle, be applied to the matrices in order to recover $f(E_0)$ at the top of the atmosphere similar to the technique described in \citep[e.g.,][]{bohm2007solar}. This is however a very challenging task due to both the ill-posed nature of the matrix inversion and the limited energy range of the measurement. (Development of a robust inversion method is in progress but not yet complete.)

To study the evolution of the particle spectra while propagating through the atmosphere, we construct different matrices $\rm{\bar A}^\sigma$ under different atmospheric column depths $\sigma$ from 1 g/cm$^2$ thickness down to the surface where the column density is about 22.5 g/cm$^2$ corresponding to pressures of 830 Pascals in a hydrostatic equilibrium state\footnote{This is the average surface pressure value over one Martian year at Gale Crater recently measured by the Rover Environmental Monitoring Station \citep[REMS,][]{gomez2012rems} onboard MSL. This is about 5-6 g/cm$^2$ greater than the column depth at the mean surface elevation, since the altitude of Gale Crater is about - 4.4 km MOLA (Mars Orbiter Laser Altimeter), but less than the column depths found in some other locations.}. 
In addition to SEP protons, which typically dominate, we have also considered primary $^4$He ions as input. The dominant secondary particle types (types $j$) include protons, $^4$He and $^3$He ions, deuterons, tritons, neutrons, gammas, electrons and positrons. 
For each primary and a given secondary type, we generate a matrix $\rm{\bar A}_{ij}^\sigma$. 
Furthermore, the particle flux reaching the surface may also produce backscattered particles, i.e., so-called albedo particles. These are produced by nuclear interactions in the regolith. 
Backscattered neutrons have been observed from orbit missions \citep[e.g.,][]{boynton2004} and {\textit{in situ}} by the DAN instrument aboard Curiosity in its ``passive'' mode \citep{jun2013}.
Since the energy spectra of upward- and downward-traveling particles are dissimilar, we have separately constructed the upward and downward directed matrices for each primary-secondary case as $\rm{\bar A}_{ij}^{\sigma-{up}}$ and $\rm{\bar A}_{ij}^{\sigma-{dn}}$ respectively. Therefore the total downward {or upward} spectra of particle type j generated by different primary particle types at the depth of $\sigma$ are: 
\begin{eqnarray}\label{eq:matrix_multiply}
  \rm{F}_j^{\sigma-{dn}}(E_j) = \sum_{i} \rm{\bar A}_{ij}^{\sigma-{dn}}(E_0, E) \cdot f_{i}(E_0); \nonumber \\
\rm{F}_j^{\sigma-{up}}(E_j) = \sum_{i} \rm{\bar A}_{ij}^{\sigma-{up}}(E_0, E) \cdot f_{i}(E_0).
\end{eqnarray} 

Panels (a) and (b) of Figure \ref{fig:PLANETOMATRIX} show the matrices of primary protons generating secondary downward and upward protons respectively. 
The atmospheric depth in this case is about 20 g/cm$^2$ {(slightly above the surface)}. Primary protons with energies less than about 150 MeV, indicated by a vertical line in (a), lack the range to reach the surface; secondary particles with up to 150 MeV energy are from primaries with higher energies. 
A similar cutoff energy for protons has also been found by \citet{gronoff2015computation}. 
Figure \ref{fig:PLANETOMATRIX}(c) and (d) show the example of primary protons generating secondary downward and upward neutrons. 

In most solar events, protons are a large majority of the primary particles reaching the top of the Martian atmosphere. 
In some SEP events, significant numbers of helium ions are accelerated, and (energy-dependent) $^4$He/$^1$H flux ratios from a few percent up to 10\% have been observed \citep{bertsch1972, benck2016}. A ratio as large as 10\% may be considered a reasonable upper limit for the ratio of time- and energy-integrated fluxes \citep{torsti1995}. 
Fig. \ref{fig:PLANETOMATRIX_alpha} shows primary Helium ions induced secondary (a) $^4$He and (b) protons near the surface of Mars at a depth of 21 g/cm$^2$. 
Since $^4$He ions obey the same range-energy relationship as protons, the $^4$He-$^4$He matrix, like the $^1$H matrix, shows a cutoff energy for incoming particles at about 150 MeV/nuc. 
The diagonal line shows the primaries which reach this depth without losing energy in the atmosphere. Very few high energy $^4$He secondaries (larger than 2 GeV/nuc) have been generated in the atmosphere. 
However, many secondary protons are generated by primary $^4$He particles as shown in (b). 
  
Based on the matrices of primary proton and $^4$He induced secondaries, we have modeled the surface spectra and radiation environment induced by primary GCRs and SEPs. 
We have ignored heavier primary ions since they contribute only $\sim$ 1\% of the GCR flux \citep{simpson1983} and even less of the SEPs. {It is however important to note that high-Z particles may interact with the atmosphere and generate secondaries which still contribute to the surface radiation exposure \citep{dartnell2007modelling, matthia2016martian}. To construct all matrices for high-Z particles paired with each type of secondaries is highly computational and is beyond the scope of the current work. As we are more interested in the application of the PLANETOMATRIX approach to modeling SEPs, we consider the construction of matrices based on primary protons and $^4$He ions to be sufficient.}

\section{Radiation dose rates}\label{sec:calculate_dose}
Radiation dose rate is a key quantity used to evaluate the energetic particle environment. Both charged and neutral particles deposit energy while going through target materials such as skin, bones and internal organs. 
Dose is defined as the energy deposited by radiation per unit mass, integrated over time, with a unit of J/kg (or Gy). The dose rate in space is often expressed in units of \textmu Gy/day.
Dose rate is one of the essential factors to be considered for future crewed missions to deep space and to Mars. 
It is therefore very important to measure and model the GCR- and SEP-induced dose rate in the interplanetary (IP) space and on the surface of Mars.
 
For any given particle spectrum, the radiation dose rate can be calculated by the following logic \citep[e.g.,][]{guo2015modeling}:
\begin{eqnarray}\label{eq:dose_def}
D = \sum\limits_{j} \sum\limits_{area}^{} {\iint \limits_{0, 1}^{E 10^6} \lambda_j(E, \epsilon) F_j(E) dE d\epsilon}/{m}, 
\end{eqnarray}
where $j$ is the particle type, $F_j(E)$ (in the unit of counts/MeV/sec/cm$^2$/sr) is the particle spectrum, $m$ {(kg)} is the mass of the material (biological bodies or detectors) and $\epsilon$ is the energy deposited by the particle in the material{ which cannot exceed the total particle energy E. The minimum and maximum energies of particles inducing dose are bounded by the energy range used in the PLANETOMATRIX which are 1 and 10$^6$ MeV respectively.}

This energy transfer process, included as a yield factor, $\lambda_j(E,\epsilon)$, can be accurately estimated using either the Bethe-Bloch equation \citep{bethe1932bremsformel} (for charged particle ionization energy loss in an infinite volume) or with more sophisticated Monte Carlo models such as GEANT4 \citep{matthia2016martian} accounting for the probability distribution of $\epsilon$ in finite volumes as used in this study. Finally $D$ is the corresponding dose rate integrated over the entire collecting volume and all the detected particle species, per unit time, with units of MeV/kg/sec (sometimes expressed as \textmu Gy/day). 

The dose rate on the surface of Mars is - apart from a negligible natural background -
mainly determined by the GCR fluxes of both primaries and secondaries during solar quiet times and it may be enhanced  significantly during SEP events. As the interactions of particles through the atmosphere depend on the particle type, energy, and the depth of the atmosphere, we model SEP-induced spectra with a variety of spectra and a range of elevations on Mars. The resulting induced dose rates can be compared with radiation dose during solar quiet times.

\section{Interplanetary GCR and the induced spectra on the surface of Mars}\label{sec:GCR}
\begin{figure}[htb!]
\centering
\includegraphics[scale=0.75]{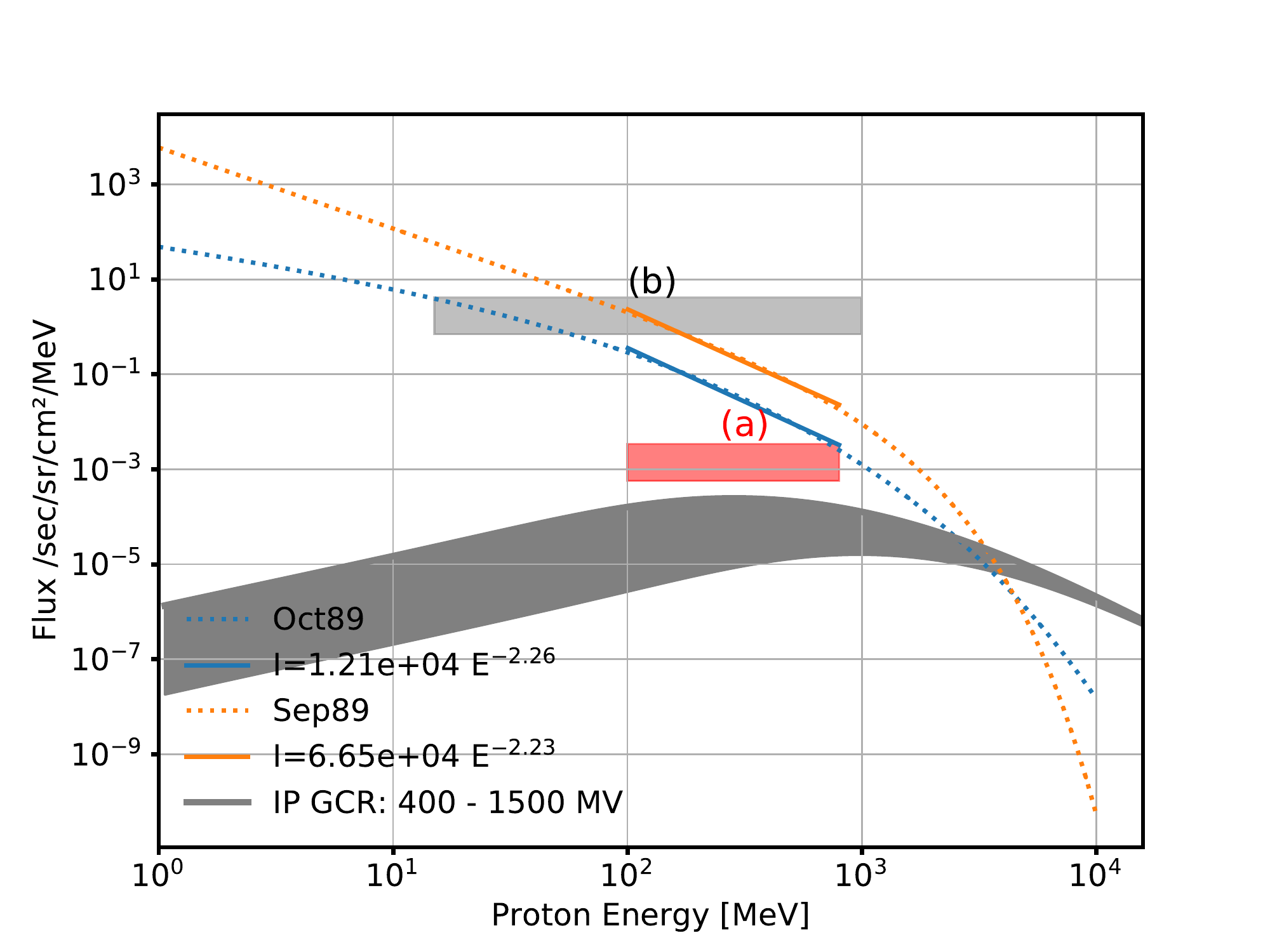}
\caption{Proton spectra of the Oct89 (blue) and Sep89 (orange) events. The gray area marks the interplanetary GCR proton spectra range when solar modulation potential varies between 400 and 1500 MV. Parts of the SEP event spectra (100-800 MeV range) are fitted with a power-law with the fit parameters shown as legends. 
The 100-800 MeV range marked by a red highlighted area is used as the energy range for the primary spectra in our case (a) study while the 15-1000 MeV range (gray area) of the spectra represents the energy range for case (b) study. The case studies are shown in Table \ref{table:3events_sum} and discussed after Section \ref{sec:EPHIN}. }\label{fig:IPspectra}
\end{figure}
The GCRs are modulated by solar activity: during solar maximum the increased solar and heliospheric magnetic fields are relatively efficient at preventing lower-energy GCRs from entering the inner heliosphere {\citep[e.g.,][]{heber2007, wibberenz2002simple}} compared to solar minimum {when the interplanetary magnetic field strength is reduced \citep{goelzer2013analysis, smith2013decline, connick2011interplanetary}}. 
That is, the GCR flux is most intense during solar minimum {\citep[e.g.,][]{mewaldt2010record, schwadron_lunar_2012}}.

In order to compare the SEP spectra and induced dose rates with those during solar quiet periods, we have employed the 2010 version of the Badwahr-O'Neill model \citep[BON10,][]{badhwar2010} to estimate GCR proton and $^4$He spectra under different modulation potentials, $\Phi$. The corresponding secondary spectra on the surface of Mars are obtained following Eq. \ref{eq:matrix_multiply}. 
Fig.\ref{fig:IPspectra} shows the GCR proton flux between {extreme modulation conditions} in a gray area. The lower-energy end of the spectra spans nearly two orders of magnitude as the modulation potential varies from 1500 MV (solar maximum) to 400 MV (solar minimum).
The long-term solar modulation of $^4$He ions is also shown in Fig. \ref{fig:sep89} (b) in gray shaded areas. 
The secondary spectra on the surface of Mars under different modulation potentials are shown in pink shaded areas through Figs. \ref{fig:oct89} to \ref{fig:sep89}.
In each panel, the surface dose rates (calculated following Eq. \ref{eq:dose_def}) are shown in the legends on the right side. 
For instance, Fig.\ref{fig:oct89}(a) shows the GCR proton dose rate as 25.6 \textmu Gy/day at $\Phi$ = 1500 MV and 171.5 \textmu Gy/day at $\Phi$ = 400 MV. 
The GCR-induced surface downward proton has a dose rate value from 18.7 to 83.6 \textmu Gy/day during solar quiet periods.
  
The figure also shows that the surface GCR spectra and dose rates are much less affected by modulation than they are in interplanetary space.
This is because the Martian atmosphere filters out lower-energy primary particles which are most affected by solar modulation. 
This effect has been supported by measurements on the surface of Mars compared to those in deep space, and the correlation between dose rate and solar modulation potential is (as expected) found to be smaller on the surface than in a spacecraft in deep space \citep{guo2015cruise, guo2015modeling}. 

\section{1989 Autumn events on the surface of Mars}\label{sec:Storia}
To simulate the large variability and effects of extreme SEP events, two historic SEP events with different spectral shape, spectral hardness, and integral proton fluence were chosen. 
The October 22nd 1989 (Oct89) event spectrum has been reconstructed using a Weibull distribution following \citet{xapsos2000characterizing} as shown Figs. \ref{fig:IPspectra} and \ref{fig:oct89}. 
For the September 1989 (Sep89) event spectrum, we use that derived by \citet{duldig1998extended} from ground level Earth neutron monitors with different rigidity cutoffs; the spectrum is shown in Figs. \ref{fig:IPspectra} and \ref{fig:sep89}.
Spectra of both events has been constructed upto 10 GeV and they have shown a sharp declining shape above $\sim$ GeV with a flux rate smaller than the ambient GCR proton spectra as shown in Fig. \ref{fig:IPspectra} and therefore we consider the contribution by SEPs with energies above $\sim 10$ GeV to the enhancement of surface dose rate in comparison to solar quiet time to be insignificant. 
In fact our calculations and comparisons considering different parts of the primary spectra and the induced exposure have demonstrated that primary protons above 1 GeV contribute to less than $\sim$ 4 percent of the surface dose rate (shown in Table \ref{table:3events_sum} with more discussions later).
In order to compare the properties of these events with power-law fitted spectra in the next section, we have also used single power-law fits to these spectra and the resulting fit parameters are shown in Fig. \ref{fig:IPspectra}. 

\subsection{Oct89 event}\label{sec:oct89}
\begin{figure}[htb!]
\centering
\begin{tabular}{cc}
\subfloat[downward proton] {\includegraphics[trim=20 25 8 63,clip, scale=0.30]{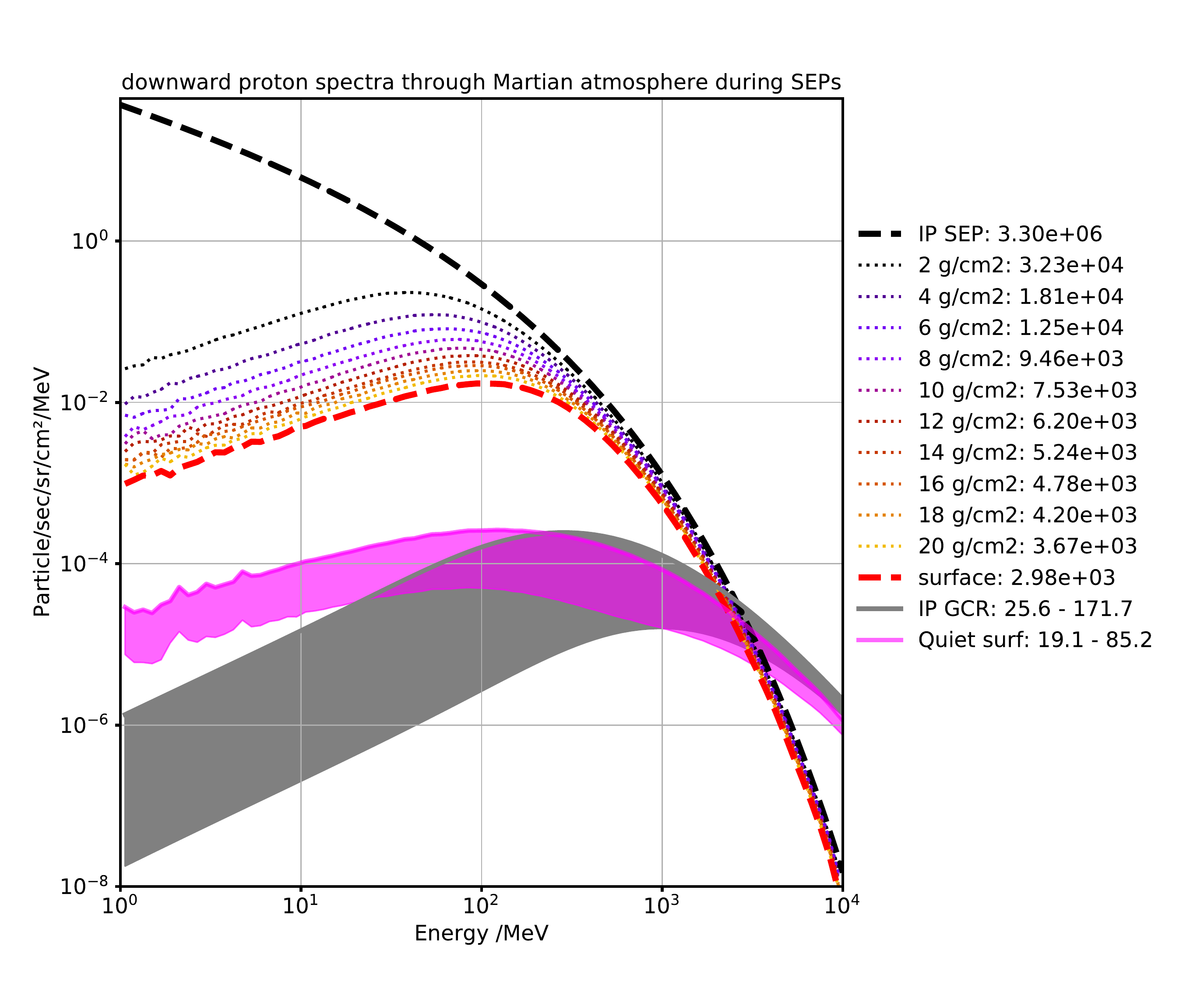} } & 
\subfloat[upward proton] {\includegraphics[trim=20 25 8 63,clip, scale=0.30]{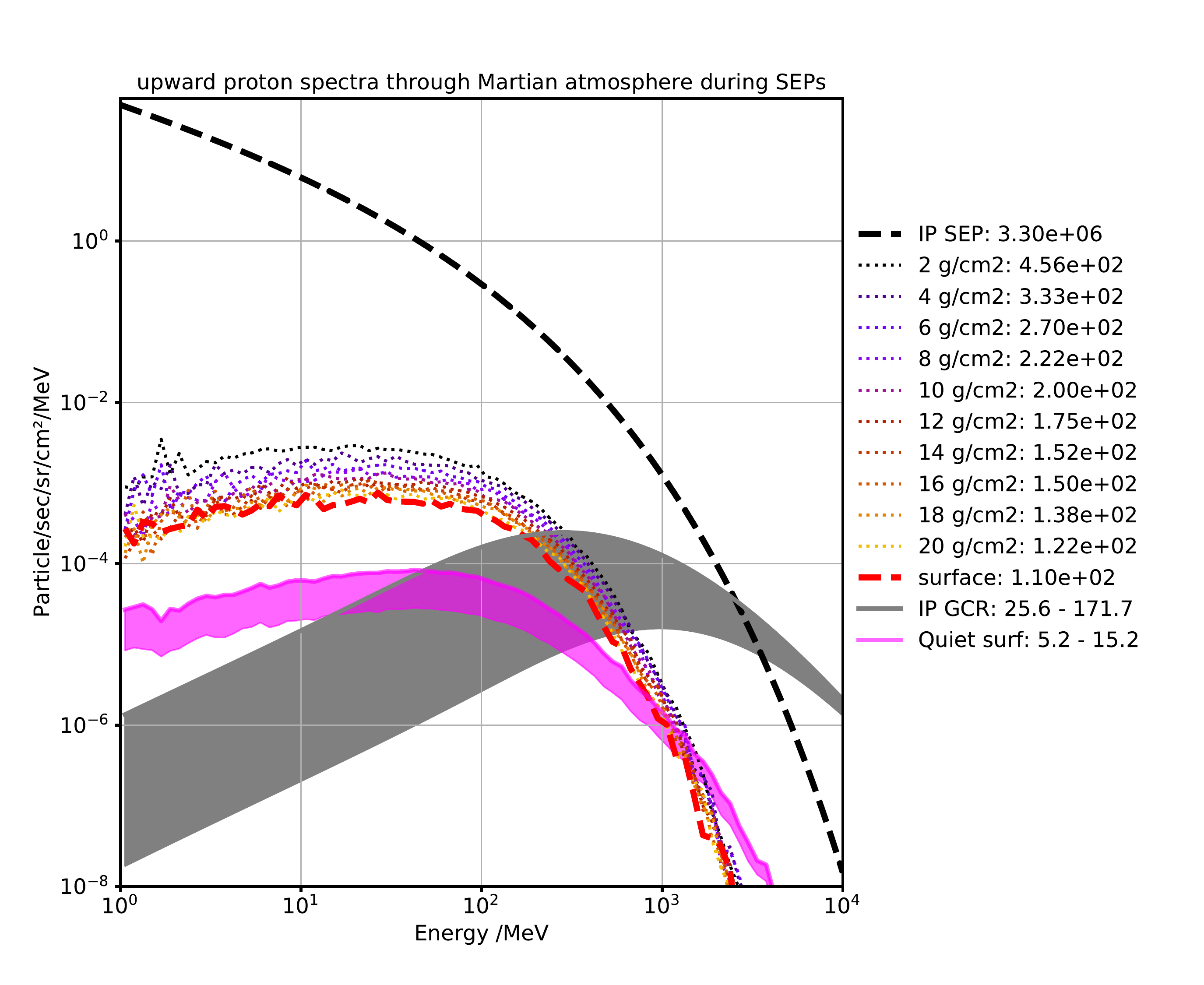} } \\
\subfloat[downward neutron] {\includegraphics[trim=20 25 8 63,clip,  scale=0.30]{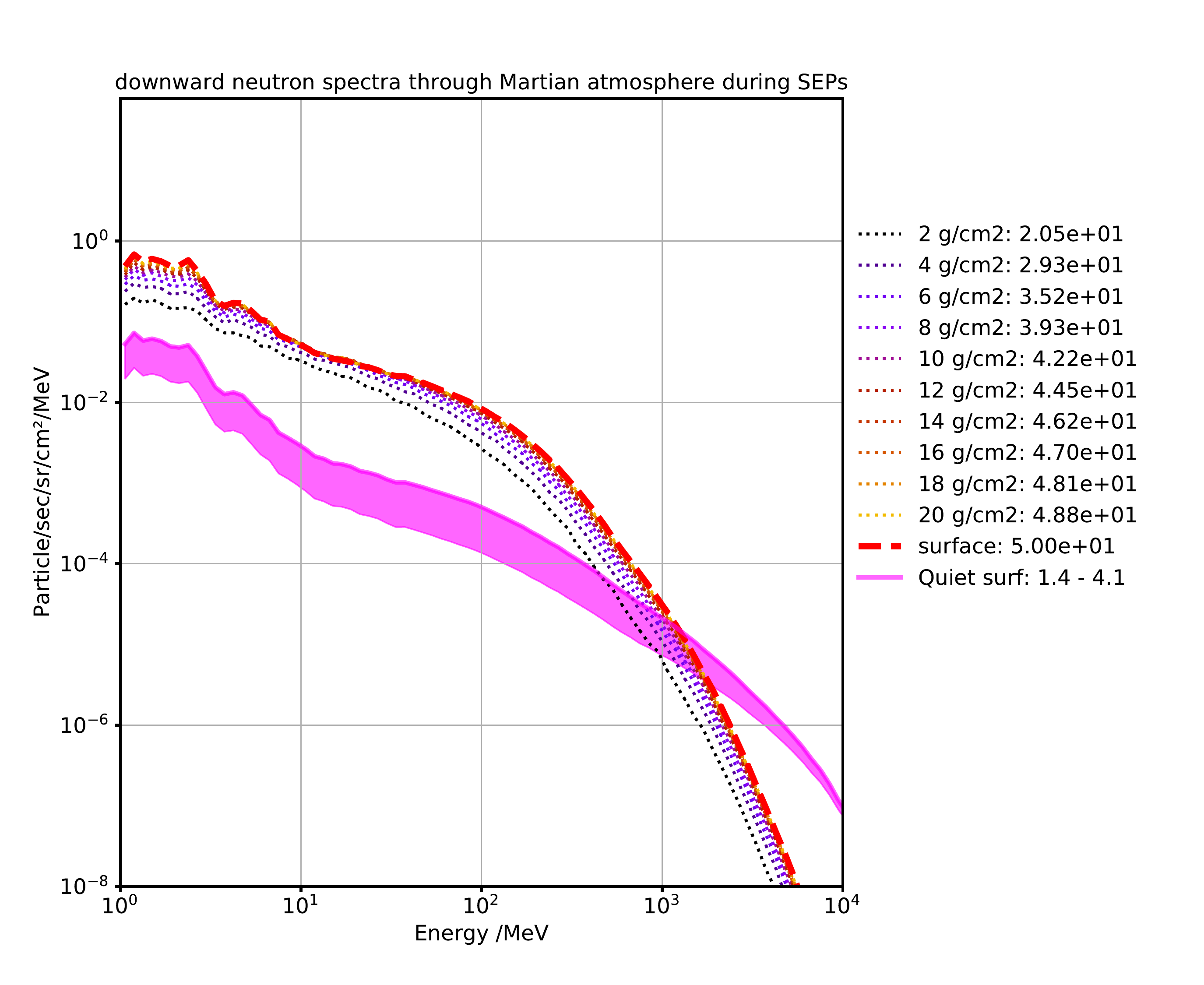} } & 
\subfloat[upward neutron] {\includegraphics[trim=20 25 8 63,clip,  scale=0.30]{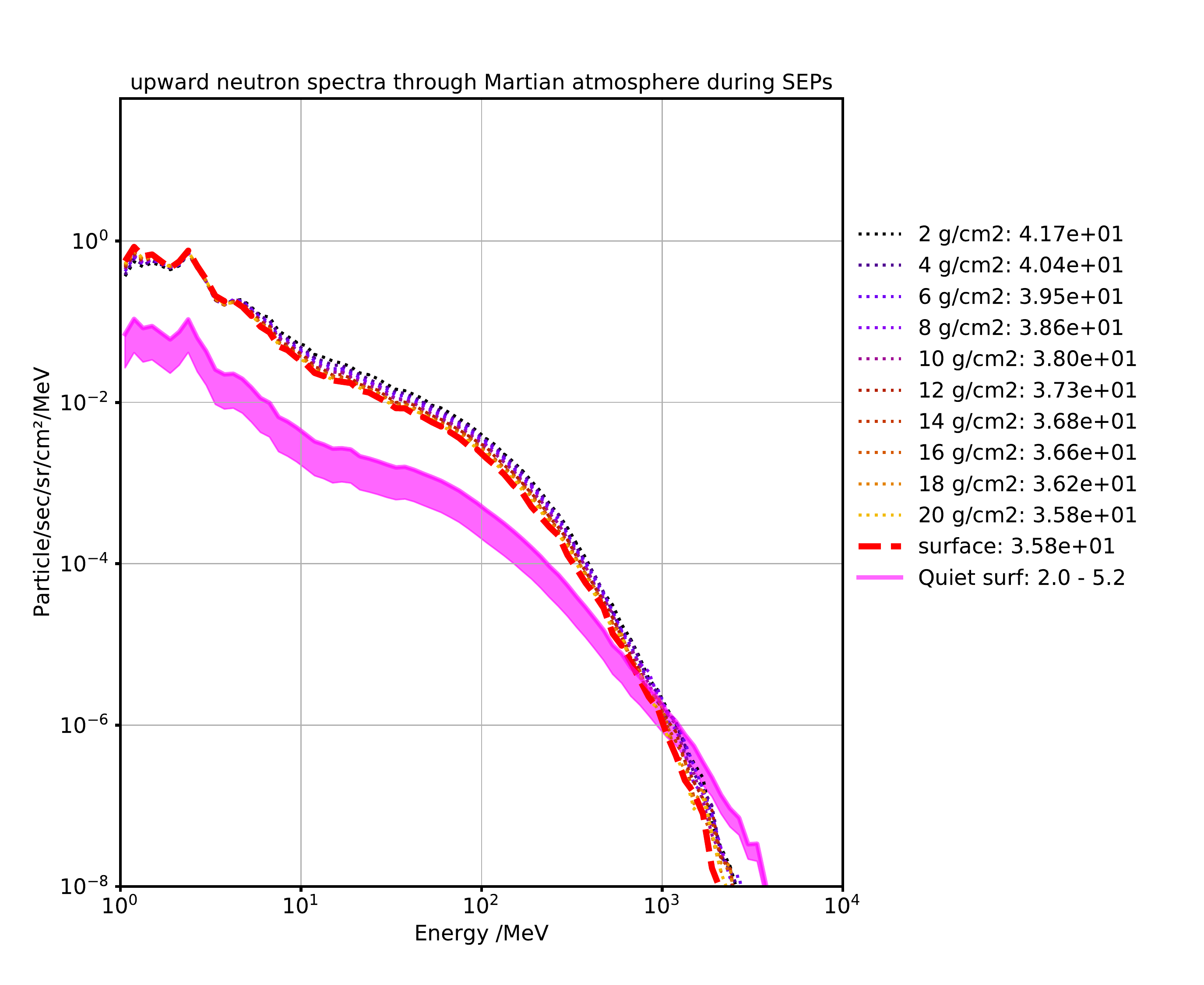} }
\end{tabular}
\caption{Particle spectra (x-axes: energy in MeV; y-axes: particles/s/sr/cm$^2$/MeV) as well as the corresponding dose rate (\textmu Gy/day, shown in legends) induced by GCRs (shaded region: gray for IP and magenta for surface) and SEPs (dashed lines) from {the Oct89 event} through the Martian atmosphere. See Section \ref{sec:oct89} for more details. }\label{fig:oct89}
\end{figure}

Fig. \ref{fig:oct89} shows the energetic particle spectra and dose rates induced by the primary proton flux associated with the Oct89 event for various atmospheric depths on Mars. 
Left (right) panels show downward (upward) spectra of protons (top) and neutrons (bottom).
The primary SEP spectra are marked by black dashed lines in each panel, and the induced secondary particle spectra are obtained by multiplying the primary SEP spectra with the corresponding PLANETOMATRIX $\rm{\bar A}_{ij}^\sigma$ at different depths $\sigma$.
For instance, Fig. \ref{fig:oct89} (a) and (b) show respectively the downward and upward proton spectra at $\sigma$ of 2, 4, 6, 8, 10, 12, 14, 16, 18 and 20 g/cm$^2$ in color-dotted lines. 
The surface spectra predicted at Gale Crater (the MSL landing site, with average atmospheric column depth of 22.5 g/cm$^2$) are plotted in thick red dashed lines. 
The proton fluxes, both upward and downward, decrease as the depth increases, particularly at lower energy. This is because the atmosphere stops a large share of the incoming protons; secondary production repopulates this part of the spectrum, but at levels far below the incident flux.

Using our modeled spectra at different depths of the atmosphere as well as on the surface of Mars (red dashed lines), we have also derived the corresponding upward or downward dose rate over a geometric angle of $2\pi$. This is the appropriate normalization since the spectra are averaged over only half the hemispheric angle. 
The dose rate values for each spectra at different depths are recorded in the legends on the right side of each panel. 
For instance, the surface downward proton dose rate of the Oct89 event spectrum is 2.98 $\times 10^3$ \textmu Gy/day, which is $\sim 10^2$ times larger than the downward proton dose rate during quiet time (19.1-85.2 \textmu Gy/day) shown as magenta region in the figure. 
The surface upward proton dose rate is 110 \textmu Gy/day for the event, considerably larger than the 5.2-15.2 \textmu Gy/day predicted during quiet time. 

To calculate the unshielded deep space dose rate, we integrate over $4\pi$ steradians, implicitly assuming isotropic fluxes for both GCRs and SEPs.
The total dose rate induced by primary GCR protons in interplanetary space ranges between 25.6 - 171.7 \textmu Gy/day for $\Phi$ between 1500 and 400 MV and the GCR spectra are shown in gray areas in the figure. 
The total dose rate induced by the Oct89 SEPs in deep space over a $4\pi$ geometric angle is estimated to be about 3.30 $\times 10^6$ \textmu Gy/day, also shown in Table \ref{table:3events_sum}, which is more than $10^4$ times higher than solar quiet time.
This enhancement ratio during the event compared to quiet time in deep space ($\sim 10^4$) is much larger than that on the surface ($\sim 10^2$). 
This is mainly because the atmosphere stops most of the SEPs, especially the low-energy ones, which contribute greatly to the unshielded deep space dose but little to the surface dose.

Fig. \ref{fig:oct89}(c) and (d) also show the proton induced secondary downward and upward neutrons respectively.
The proton flux decreases as atmospheric depth increases while most other secondary (e.g., downward electrons and neutrons) fluxes increase as the column depth increases since their fluxes build up as the SEPs penetrate into the atmosphere.
The upward neutron flux is mostly produced by protons going upward though the atmosphere, and as a result it follows the opposite trend: at points deeper atmosphere, the intensity is slightly smaller. 
For all SEP secondaries, the spectral shapes differ from GCR-induced secondary spectra in that they have fewer highly energetic particles above a few GeV. 

\subsection{Sep89 event} \label{sec:sep89}
\begin{figure}[htb!]
\centering
\begin{tabular}{cc}
\subfloat[downward proton] {\includegraphics[trim=20 25 8 63,clip, scale=0.30]{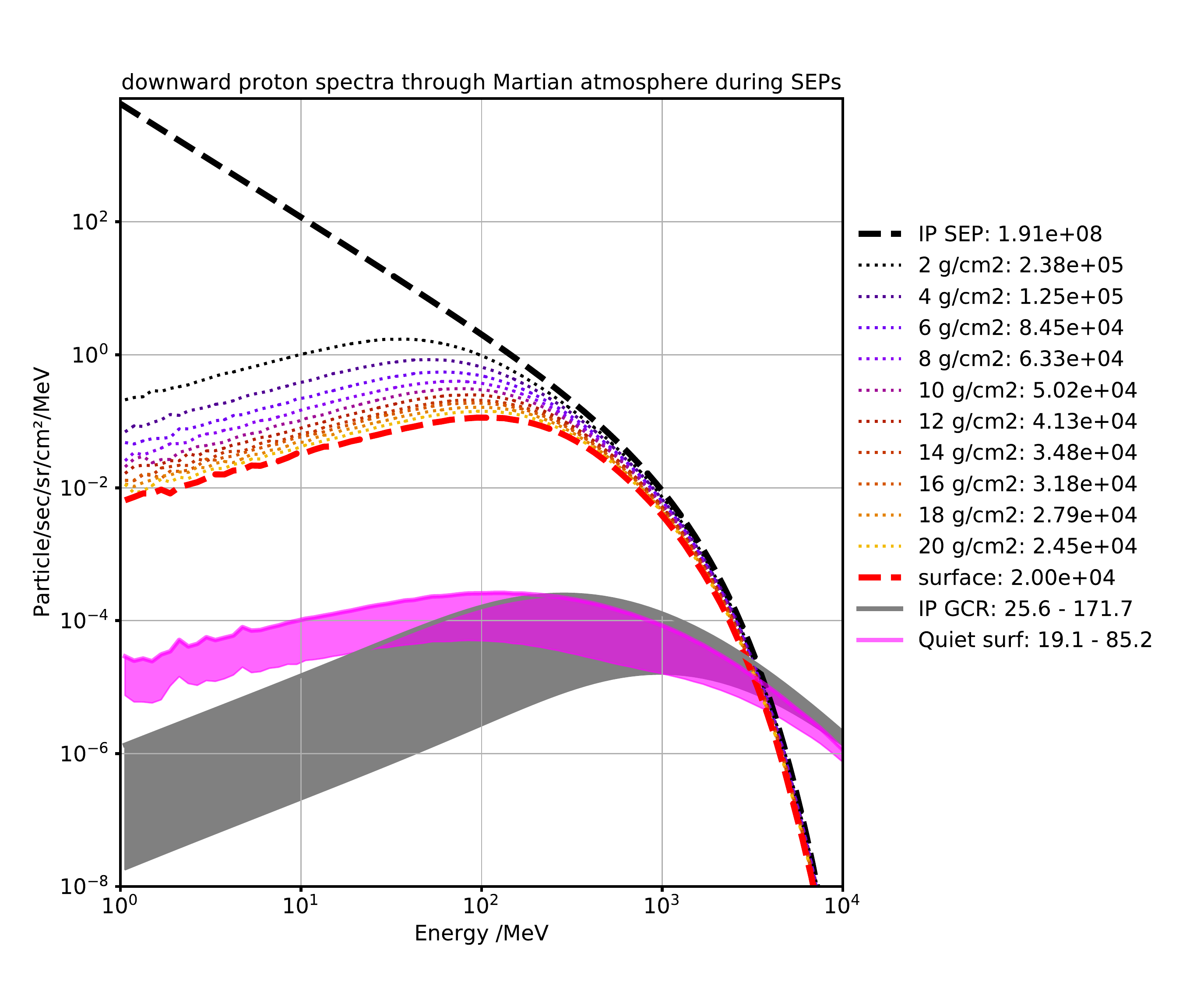} } & 
\subfloat[downward $^4$He] {\includegraphics[trim=20 25 8 63,clip, scale=0.30]{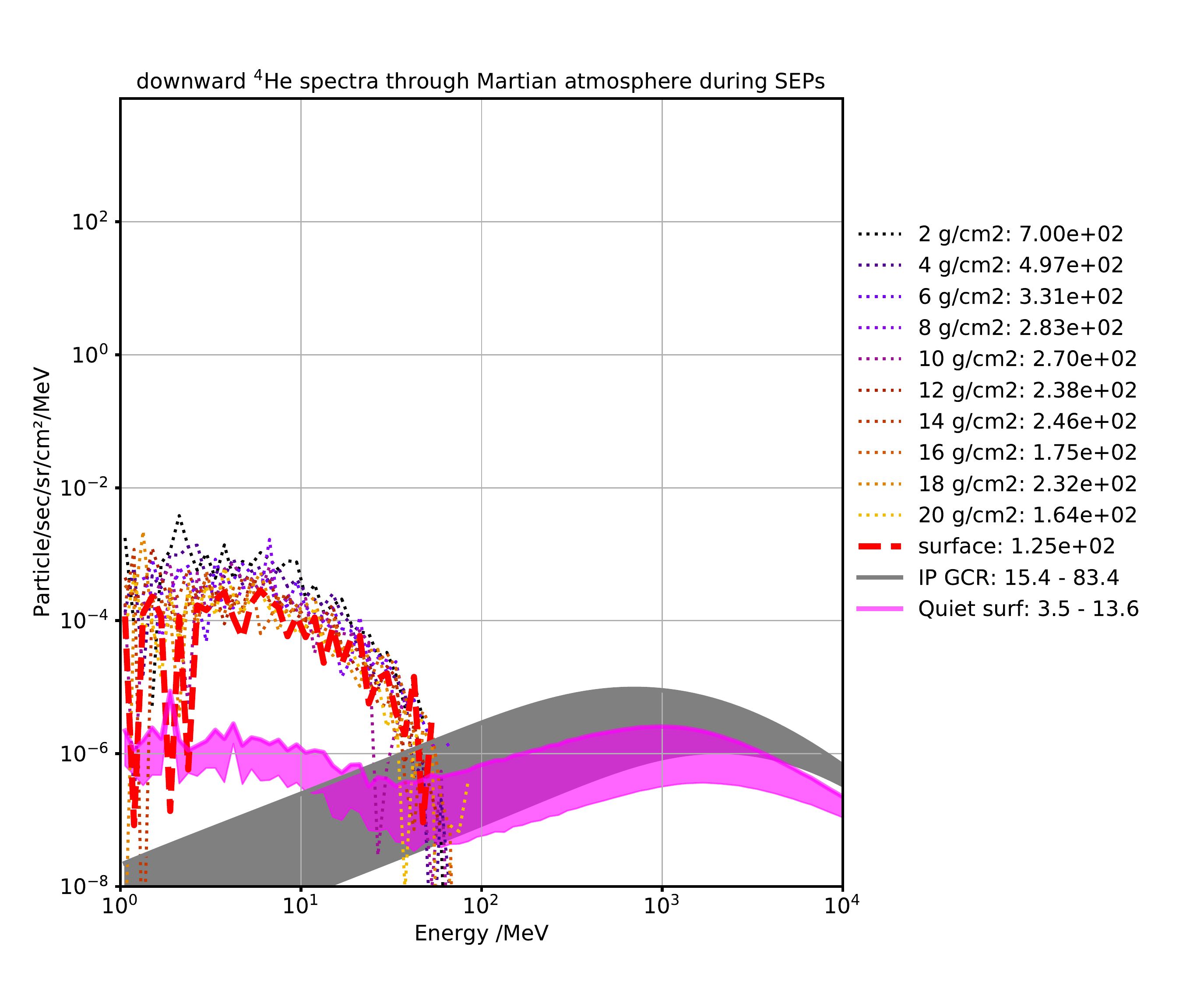} } \\
\subfloat[downward triton] {\includegraphics[trim=20 25 8 63,clip, scale=0.30]{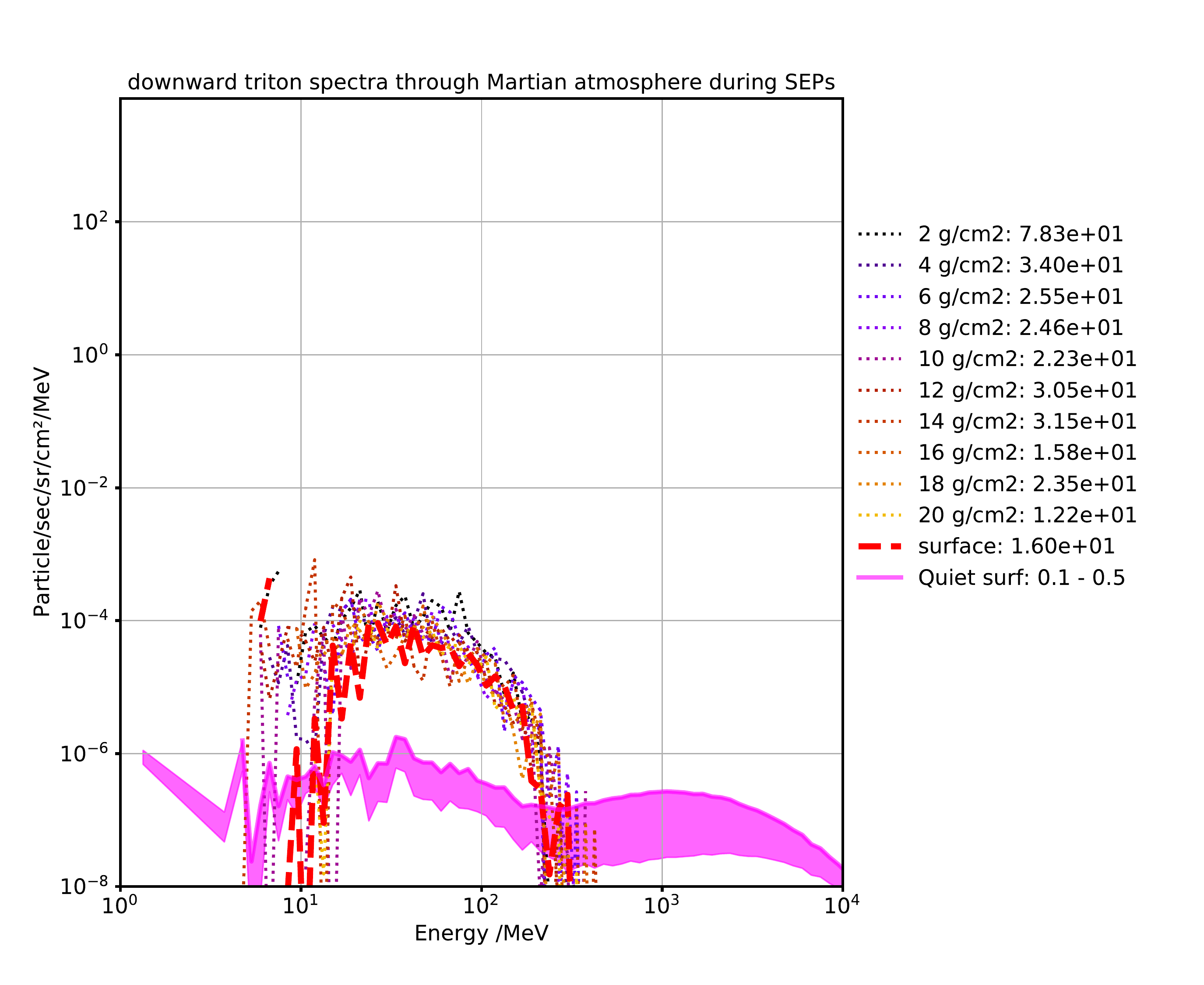} } & 
\subfloat[downward deuteron] {\includegraphics[trim=20 25 8 63,clip, scale=0.30]{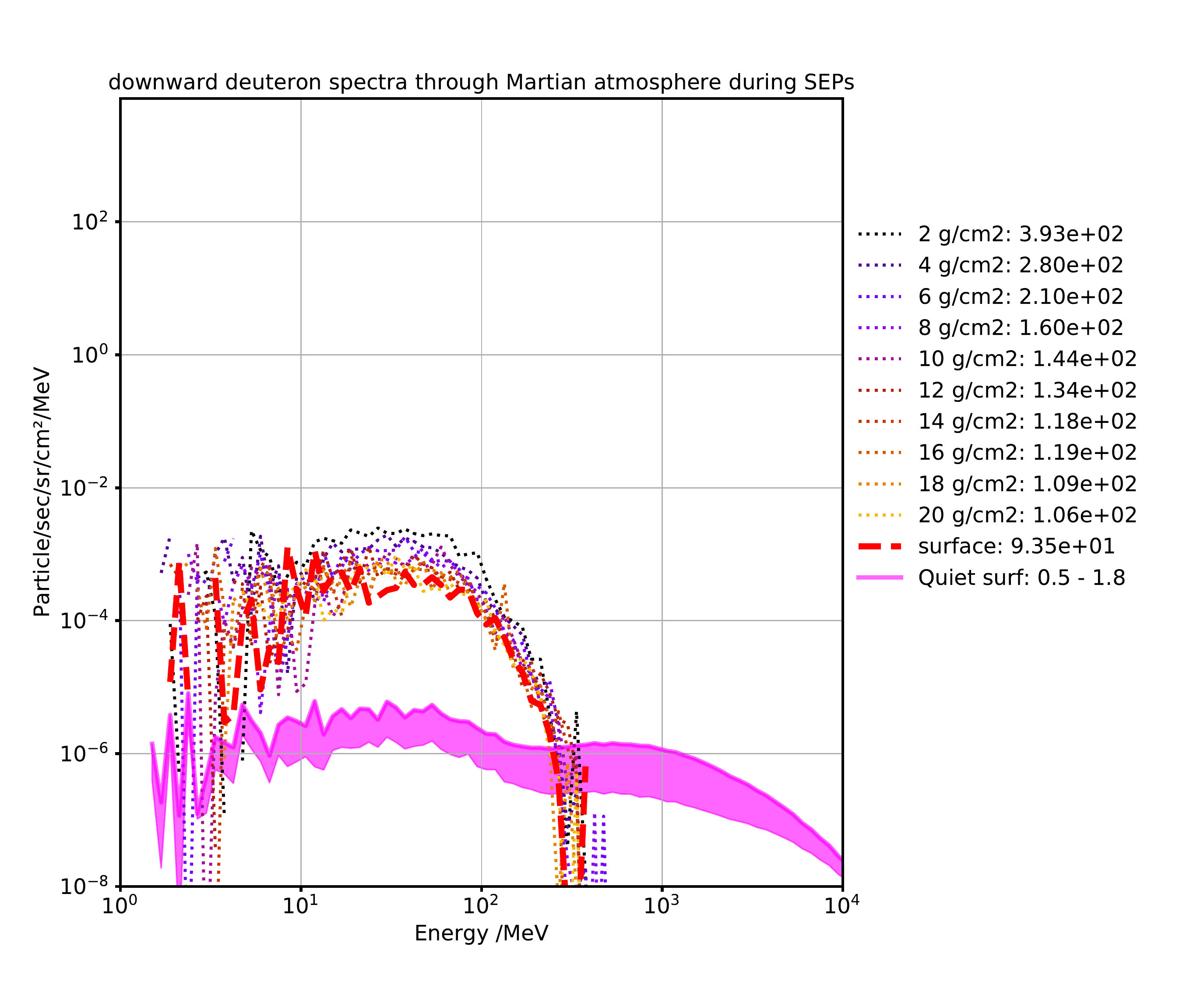} } \\
\end{tabular}
\caption{Particle spectra (x-axes: energy in MeV; y-axes: particles/s/sr/cm$^2$/MeV) and the corresponding dose rate (\textmu Gy/day, shown in legends) induced by GCRs (shaded region: gray for IP and magenta for surface) and SEPs (dashed lines) from {the Sep89 event} through the Martian atmosphere. See Section \ref{sec:sep89} for more details.}\label{fig:sep89}
\end{figure}

The Sep89 event has a quasi-power-law spectra at energies up to about 1 GeV and has been well-represented by a single power-law spectral shape between 100 and 800 MeV as shown in Figs. \ref{fig:IPspectra} and \ref{fig:sep89}. 
Its spectral shape is slightly different from that for the Oct89 event with its high energy component ($\ge 1$ GeV) very soft with a sharper drop-off.
Because of the high intensity, the Sep89 event produces significant enhancements on the Martian surface for all different types of secondaries including those which are generated with much lower probabilities such as $^4$He ions, tritons and deuterons as shown in Fig.\ref{fig:sep89}. 

A summary table of the total dose rates (both upward and downward directions for the Oct89 and Sep89 events) from interplanetary deep space through the atmosphere down to the surface are shown in Table \ref{table:3events_sum}.
During the Oct89 event, the total surface dose rate including both downward and upward secondaries is about 3.30 $\times 10^3$ \textmu Gy/day, only about 0.1 \% times compared to the SEP dose rate in unshielded deep space. 
For the Sep89 event, this ratio is even smaller, due to the effect shielding of the Martian atmosphere against low-energy primaries which contribute largely to the deep space dose rate. 

We have tried to compare our results of the surface dose rate to those from \citet{gronoff2015computation} for the these two events. They have obtained the surface dose rates of 9.4$\times 10^3$ and 6.5 $\times 10^4$ \textmu Gy/day for Oct89 and Sep89 events respectively. These values are approximately 3 times of what we obtained here. Careful investigation has revealed that they have used the input SEP spectra 4 times of those provided in the parameters and equations \citep{norman2014influence, lovell1998extended} resulting in a higher surface dose rates compared to ours. Their particle transport model is however still valid as (a) they have reached similar results between two different models (HZETRN and PLANETOCOSMICS) and (b) we have derived similar dose rates when using the SEP input sepctra they used. 

{Furthermore, the choice of different low-cutoff energies for calculating dose may result in different results since the linear energy transfer $dE/dx$ function which goes as $ \sim 1/v^2$ ($v$ is the proton velocity) gives very large values for low-energy particles. The influence of this cutoff energy is more significant for SEP spectra in deep space where low-energy particles are much more abundant and thus it should always be clearly notified in dose rate calculations. 
A cutoff energy of 1 MeV has been used throughout this study (also see Section \ref{sec:calculate_dose}) and this is actually rather low as protons with energies smaller than e.g., $\sim 5$ MeV would hardly penetrate through a 0.5 mm water slab.}

  \begin{deluxetable}{c|rrr|rrr}
  \tabletypesize{\scriptsize}
  \tablecaption{The {upward, downward and} total dose rates [\textmu Gy/day] in deep space and at different atmospheric depths of Mars for the Oct89 and Sep89 SEP events. 
 {Results of three different case studies are shown here: (f) full primary proton spectra with protons larger than 1 MeV used as the input spectra through the PLANETOMATRIX approach, (a) only primary protons from 100 to 800 MeV and (b) primary protons from 15 to 1000 MeV.}
  The 'ratio' row shows the ratio of the surface dose rate to the deep space dose rate. 
  For instance, (fs)/(fd) ratio is the full-spectra induced surface dose rate divided by the full-spectra induced deep space dose rate; (as)/(fs) ratio is the case (a) surface dose rate divided by the full-spectra surface dose rate.  \label{table:3events_sum}}
  \tablewidth{0pt}
 \startdata
 \tableline
  & Oct89 upward & downward & total & Sep89 upward & downward & total \\
 \tableline
  (f) primary protons &&&&&&\\
	$>$ 1 MeV &&&&&&\\
  \hdashline
  deep space (fd)& 1.65 $\times 10^6$ & 1.65 $\times 10^6$ & 3.30 $\times 10^6$ & 9.95$\times 10^7$ & 9.95$\times 10^7$ & 1.91 $\times 10^8$\\
  2 g/cm$^2$  & 5.92$\times 10^2$ & 3.25$\times 10^4$ & 3.31$\times 10^4$ & 4.07$\times 10^3$ & 2.40 $\times 10^5$ & 2.44 $\times 10^5$\\
  4 g/cm$^2$  & 4.41$\times 10^2$ & 1.83$\times 10^4$ & 1.87$\times 10^4$ & 2.98$\times 10^3$ & 1.26 $\times 10^5$ & 1.29 $\times 10^5$\\
  6 g/cm$^2$  & 3.65$\times 10^2$ & 1.27$\times 10^4$ & 1.30$\times 10^4$ & 2.45$\times 10^3$ & 8.55 $\times 10^4$ & 8.80 $\times 10^4$\\
  8 g/cm$^2$  & 3.16$\times 10^2$ & 9.60$\times 10^3$ & 9.92$\times 10^3$ & 2.13$\times 10^3$ & 6.43 $\times 10^4$ & 6.64 $\times 10^4$\\
  10 g/cm$^2$ & 3.11$\times 10^2$ & 7.68$\times 10^3$ & 7.99$\times 10^3$ & 2.09$\times 10^3$ & 5.12 $\times 10^4$ & 5.33 $\times 10^4$\\
  12 g/cm$^2$ & 2.60$\times 10^2$ & 6.34$\times 10^3$ & 6.60$\times 10^3$ & 1.76$\times 10^3$ & 4.23 $\times 10^4$ & 4.40 $\times 10^4$\\
  14 g/cm$^2$ & 2.35$\times 10^2$ & 5.39$\times 10^3$ & 5.63$\times 10^3$ & 1.59$\times 10^3$ & 3.58 $\times 10^4$ & 3.74 $\times 10^4$\\
  16 g/cm$^2$ & 2.30$\times 10^2$ & 4.92$\times 10^3$ & 5.15$\times 10^3$ & 1.56$\times 10^3$ & 3.27 $\times 10^4$ & 3.43 $\times 10^4$\\
  18 g/cm$^2$ & 2.24$\times 10^2$ & 4.35$\times 10^3$ & 4.57$\times 10^3$ & 1.51$\times 10^3$ & 2.89 $\times 10^4$ & 3.04 $\times 10^4$\\
  20 g/cm$^2$ & 2.00$\times 10^2$ & 3.80$\times 10^3$ & 4.00$\times 10^3$ & 1.36$\times 10^3$ & 2.54 $\times 10^4$ & 2.68 $\times 10^4$\\
  surface (fs) & 1.82$\times 10^2$& 3.11$\times 10^3$ & 3.30$\times 10^3$ & 1.23$\times 10^3$ & 2.09 $\times 10^4$ & 2.21 $\times 10^4$\\
  \hdashline
  (fs)/(fd) ratio & 0.01 \% & 0.19 \% & 0.1 \% & 0.001\% & 0.02 \% & 0.01 \%\\
  \tableline
  (a) primary protons &&&&&&\\
  100-800 MeV  &&&&&&\\
  \hdashline
  deep space (ad) & 1.10 $\times 10^4$ & 1.10 $\times 10^4$ & 2.19 $\times 10^4$ & 7.3 $\times 10^4$ & 7.3 $\times 10^4$ & 1.46 $\times 10^5$ \\
  surface (as) & 1.34 $\times 10^{2}$ & 2.93 $\times 10^{3}$ & 3.06 $\times 10^{3}$ &  9.16 $\times 10^{2}$ & 1.96 $\times 10^{4}$ & 2.05 $\times 10^{4}$\\
  \hdashline
  (ad)/(fd) ratio & 0.66\% & 0.66\% & 0.66\% & 0.07\% & 0.07\% & 0.07\% \\
  (as)/(fs) ratio & 73.6\% & 94.2\% & \textbf{92.7\%} & 74.5\% & 93.8\% & \textbf{92.8\%}\\
  \tableline
  (b) primary protons &&&&&&\\
  15-1000 MeV &&&&&&\\
  \hdashline
  deep space (bd) & 1.49 $\times 10^5$& 1.49 $\times 10^5$ & 2.97 $\times 10^5$ & 1.59 $\times 10^6$ & 1.59 $\times 10^6$& 3.18 $\times 10^6$\\
  surface (bs) & 1.49 $\times 10^{2}$ & 3.00 $\times 10^{3}$ & 3.15 $\times 10^{3}$ & 1.03 $\times 10^{3}$ & 2.02 $\times 10^{4}$ & 2.12 $\times 10^{4}$\\  
  \hdashline
  (bd)/(fd) ratio & 9.03\% & 9.03\% & 9.03\% & 1.60\% & 1.60\% & 1.60\% \\
  (bs)/(fs) ratio & 81.9\% & 96.5\% & \textbf{95.5\%} & 83.7\% & 96.7\% & \textbf{95.9\%} \\
 \enddata
  \end{deluxetable}

\section{Twenty years of significant events modeled on the surface of Mars}\label{sec:EPHIN}
\begin{figure}[htb!]
	\centering
    \includegraphics[trim=16 20 12 35, clip, scale=0.65] {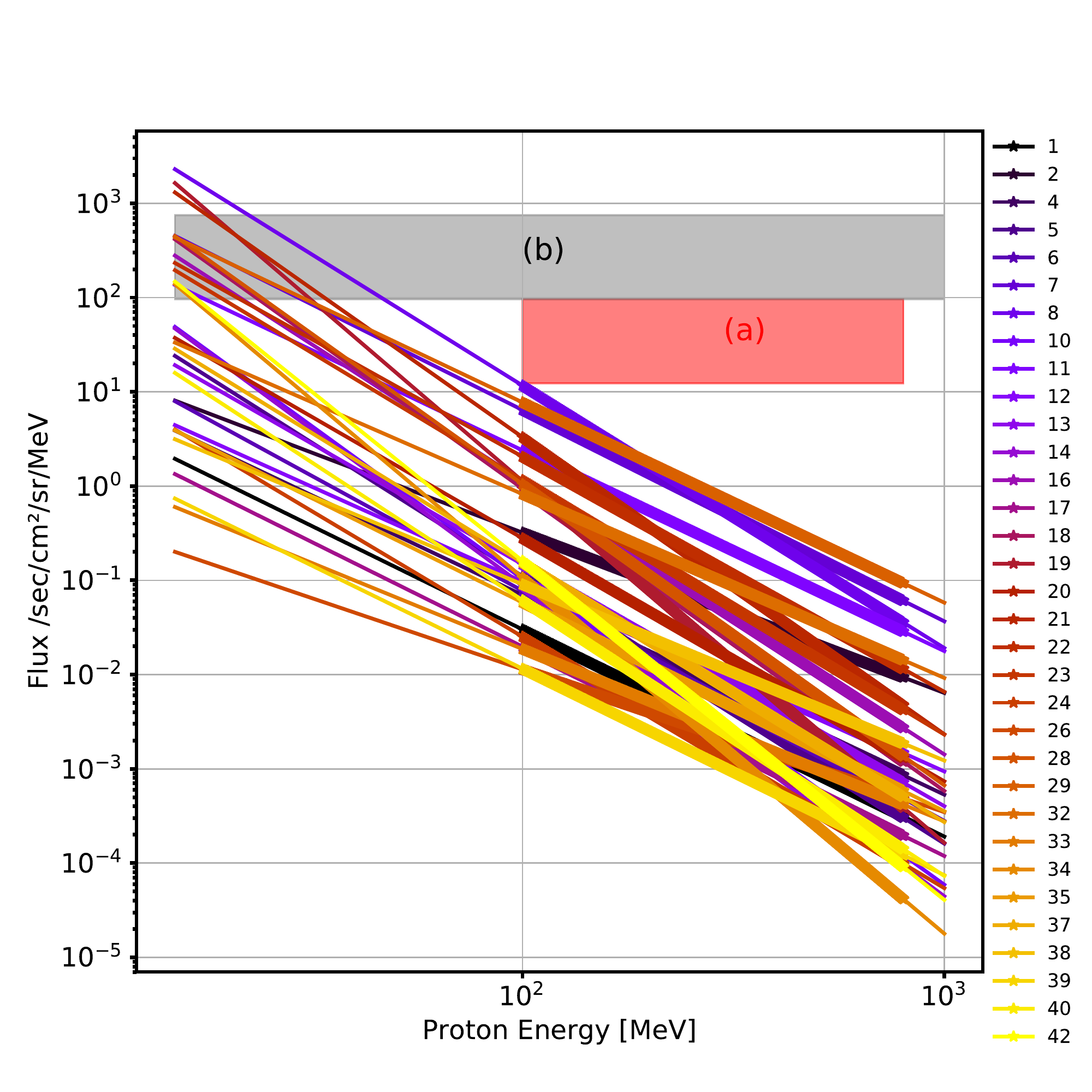}
	\caption{The fitted power-law spectra of significant SEPs detected by SOHO/EPHIN in 20 years. Different events are colored differently and the number of each spectra corresponds to the event number used in \citet{kuehl2016}. The 100-800 MeV range marked by a red highlighted area and in thick lines is used as the energy range for the primary spectra in our case (a) study while the 15-1000 MeV range (gray area) including thinner lines of the spectra represents the energy range for case (b) study. }\label{fig:EPHINspectra}
\end{figure}

\citet{kuehl2016} have studied a set of SEP events with the Electron Proton Helium Instrument (EPHIN) instrument onboard SOHO based on a newly developed optimization technique that exploits the response function of the penetrating protons through the detector sets and thus extends the usable energy range of the instrument from 5 to 50~MeV up to above 800~MeV \citep{kuehl2015}.
The studied SEP events are between December 1995 and December 2015 with protons accelerated to energies greater than 500~MeV. 
A total of 42 events has been found, including all GLEs during the SOHO age, excluding one GLE during which EPHIN had a data gap. Due to the long lifetime of the instrument, its highly efficient operation during the mission, and the fact that observations spanned different phases of the solar cycle, the total number of events is likely typical for a 20~year time period. Supporting this supposition, we note that the range of monthly-average sunspot number during this period is in line with historical records during different solar activity levels.  
For 33 of the events, the onset time is based on an energy channel covering from 100~MeV to 1~GeV; and proton spectra from 100~MeV up to 800~MeV were derived in a two-hour interval starting 30~minutes after this onset for each event. Thus the spectra may not represent the maximum intensity in every single energy bin individually, {but more likely in the very-high-energy channel which is most important for the surface radiation environment (due to the Martian atmospheric shielding)}.
A single power law function was applied to fit each spectrum according to:
\begin{eqnarray}\label{eq:power-law}
f(E_0) = I_{\epsilon_0} \cdot (E_0/\epsilon_0)^\gamma
\end{eqnarray}
where the SEP proton energy $E_0$ is in MeV, flux $f(E_0)$ and $I_0$ are in particles/sec/cm$^{2}$/sr/MeV and $I_{\epsilon_0}$ is the scaled intensity at $\epsilon_0$ MeV. 
{The fitted power-law spectra for all events used in this study is shown in Fig. \ref{fig:EPHINspectra}. }

Section \ref{sec:Storia} focused on individual historical events and the modeling of each SEP spectra in a wide energy range from 1 up to 10$^4$ MeV.
Here, we obtain an empirical correlation between deep space and Martian surface dose rates with the general properties of the SEP spectra as represented by $I_{\epsilon_0}$ and the power-law index $\gamma$.
Based on the fitted power-law parameters of the 33 events by \citet{kuehl2016}, we use PLANETOMATRIX to forward-model the power-law fitted SEP spectra $f(E_0)$, shown in Fig. \ref{fig:EPHINspectra}, through the Martian atmosphere, and obtain the induced surface secondary spectra $F_i(E)$ of different particle species.

For each SEP event, the deep space and Martian-surface dose rates have been calculated following Eq. \ref{eq:dose_def} for two ranges of the primary SEP spectra: (a) the primary proton energy range of 100-800 MeV and (b) an extended energy range of 15-1000 MeV for the primary protons. {The energy ranges of the two cases are also shown in highlighted areas in Fig. \ref{fig:EPHINspectra}.}
The two-range study is motivated by these considerations: case (a) is the trustworthy energy range for each single power-law fit spectrum \citep{kuehl2016}, while case (b) extrapolates the spectra to a wider energy range, which may yield more reliable estimates of the interplanetary dose rates contributed by events with complete energy spectra.
To avoid over-estimation of the total dose rate, we did not extrapolate the spectra to a much wider energy range since the power-law shape generally flattens out at low energies around 10-30 MeV depending on individual events \citep{band1993batse} and drops off quickly at high energies $\ge$ 1000 MeV as also shown in Fig. \ref{fig:IPspectra} for the Sep89 and Oct89 events. 
The low-energy end of 15 MeV in case (b) has also been chosen following \citet{wilson2006spacesuit} which suggested little dose contribution from protons below $\sim$20 MeV for astronauts wearing a space suit during Extra-Vehicular Activities (EVA). The interplanetary-space dose rates could therefore be considered to represent an exposure scenario in which an astronaut is doing extra-vehicular activities when an SEP event occurs.
In both cases (a) and (b), we expect the induced Martian surface dose rates to have similar values, since the atmosphere stops low energy protons (energies $\le 140$ to $\sim 160$ MeV depending on elevation). 
We also note that although protons with energies lower than this cutoff cannot reach the surface, their secondaries, especially electrons and neutrons produced in the atmosphere, can travel downward and contribute to the surface dose rates.

\begin{figure}
\centering
\begin{tabular}{cc}
\subfloat[SEP proton energy: 100-800 MeV, ${\epsilon_0}=300$ MeV] {\includegraphics[trim=10 0 500 0, clip, scale=0.5]{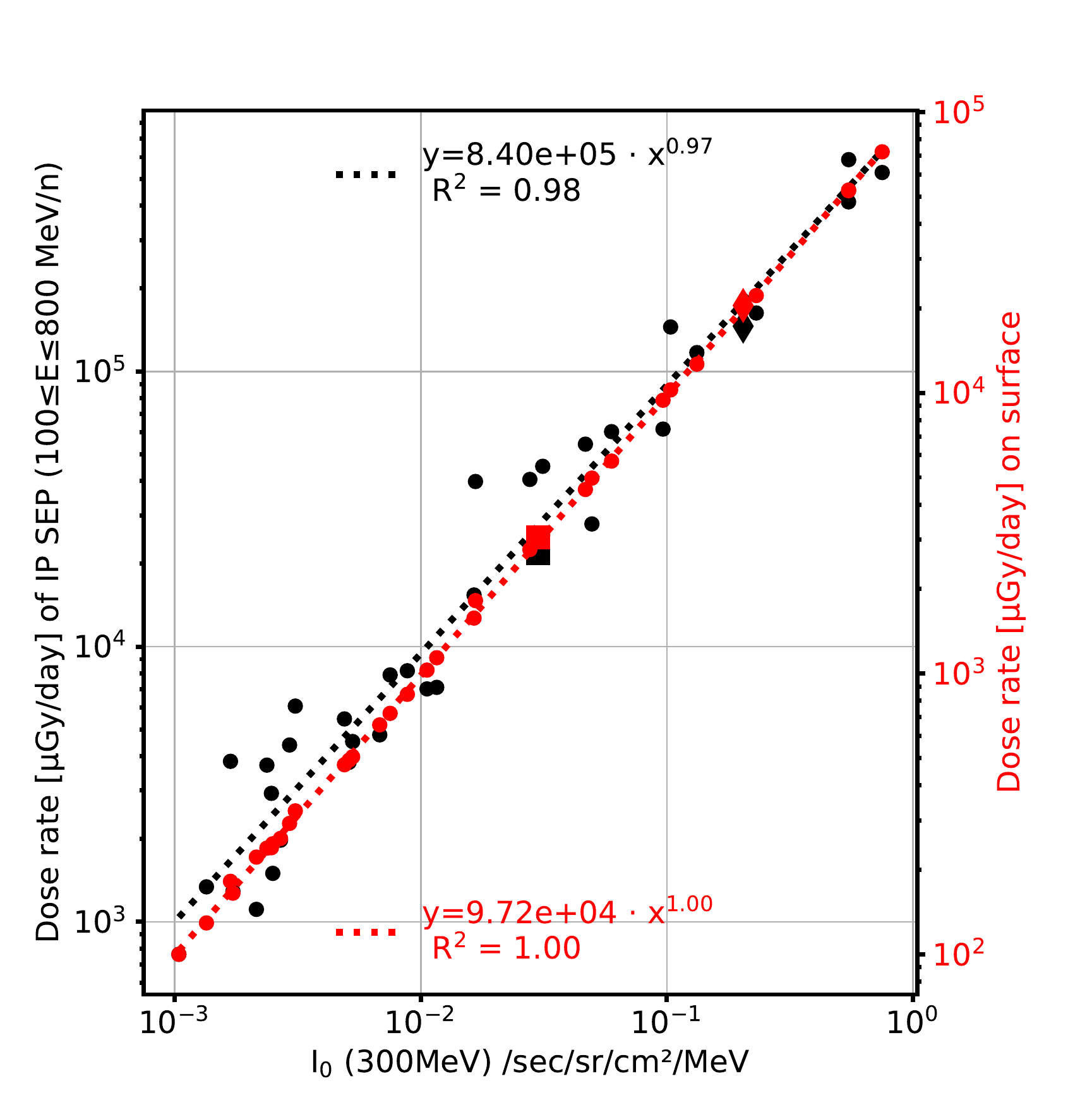}} & 
\subfloat[SEP proton power-law index $\gamma$] {\includegraphics[trim=510 0 0 0, clip, scale=0.5]{doseVS_I300-gamma_100-800MeV-proton-Water.pdf}} \\
\subfloat[SEP proton energy: 100-800 MeV, ${\epsilon_0}=200$ MeV] {\includegraphics[trim=10 0 500 0, clip, scale=0.5] {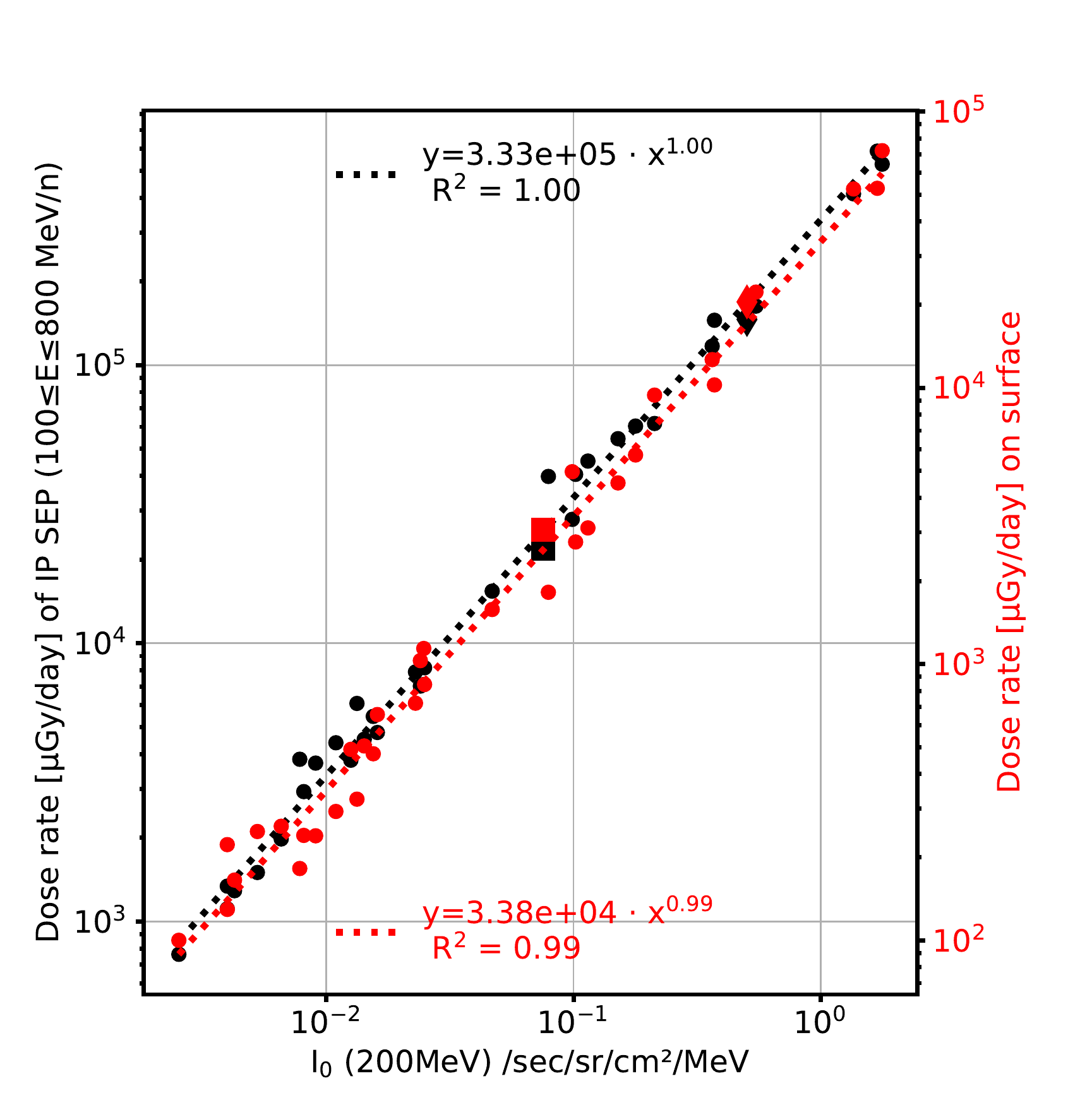}} & 
\subfloat[SEP proton integrated flux from 100 to 800 MeV] {\includegraphics[trim=0 0 510 0, clip, scale=0.5] {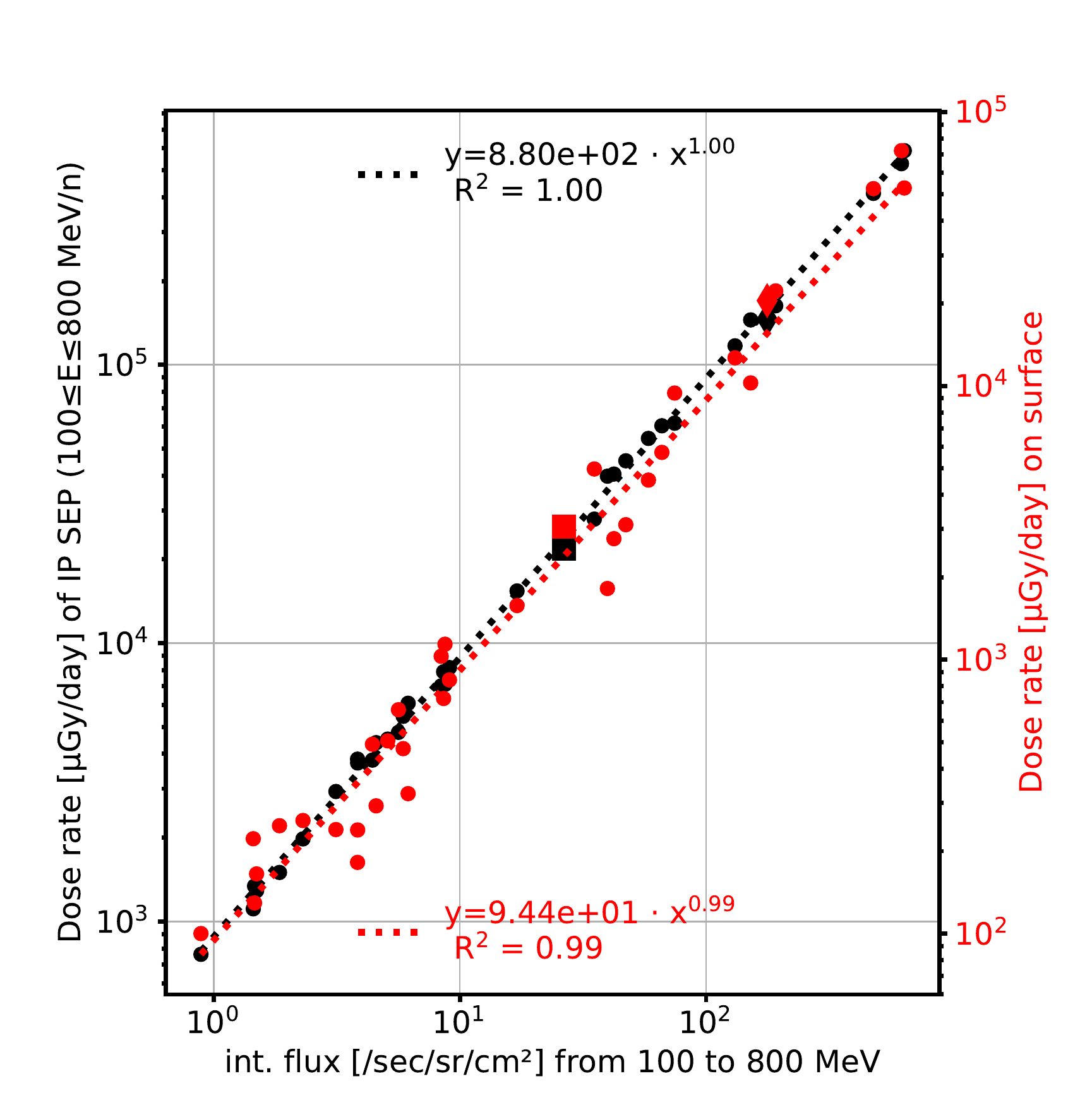}}
\end{tabular}
\caption{Case (a, primary protons from 100 to 800 MeV) results: deep space (black, left y-axes) and Martian surface (red, right y-axes) dose rate versus SEP flux $I_{\epsilon_0}$ (panels a and c) and $\gamma$ (panel b) as well as the integrated flux (panel d). 
EPHIN events are marked in circular dots; Oct89 and Sep89 events are shown in squares and diamonds respectively. More explanations of the figure are shown in text of Section \ref{sec:EPHIN}. } \label{fig:doseVSI0-gamma-casea}
\end{figure}

Fig. \ref{fig:doseVSI0-gamma-casea} summarizes the result of case (a) where 100-800 MeV primary protons were considered for each SEP spectrum with varying intensities and power-law spectral indices. 
The calculated dose rates in deep space (in black with scales on left axes) and on the Martian surface (in red and right axes) are plotted versus the SEP flux $I_{\epsilon_0}$ (panel a for $\epsilon_0=300$ MeV and c for $\epsilon_0=200$ MeV) and spectral index $\gamma$ (panel b) as well as the integrated flux (panel d) of the 100-800 MeV proton spectra. 
It is clear from panels (a) and (c) that both the deep-space and Martian surface dose rates correlate very well with $I_{\epsilon_0}$ and the fitted logarithmic linear function is plotted in dashed lines and shown as legends in the plots.
We found the best correlation (R$^2 \geq$ 1) between the Martian surface dose rates and the intensity $I_{\epsilon_0}$ at 300 MeV.
{However no clear correlation is found between dose rate and the power-law index $\gamma$. 
This is probably because the intensities of different SEPs studied here vary over more than 3 orders of magnitudes while the power-law index (ranging from -1.5 to -4) plays a minor role in determining the overall intensity of the event as shown in Fig. \ref{fig:EPHINspectra}.} 
 
\begin{figure}
\centering
\begin{tabular}{cc}
\subfloat[SEP proton energy: 15-1000 MeV, ${\epsilon_0}=300$ MeV] {\includegraphics[trim=10 0 500 0, clip, scale=0.5] {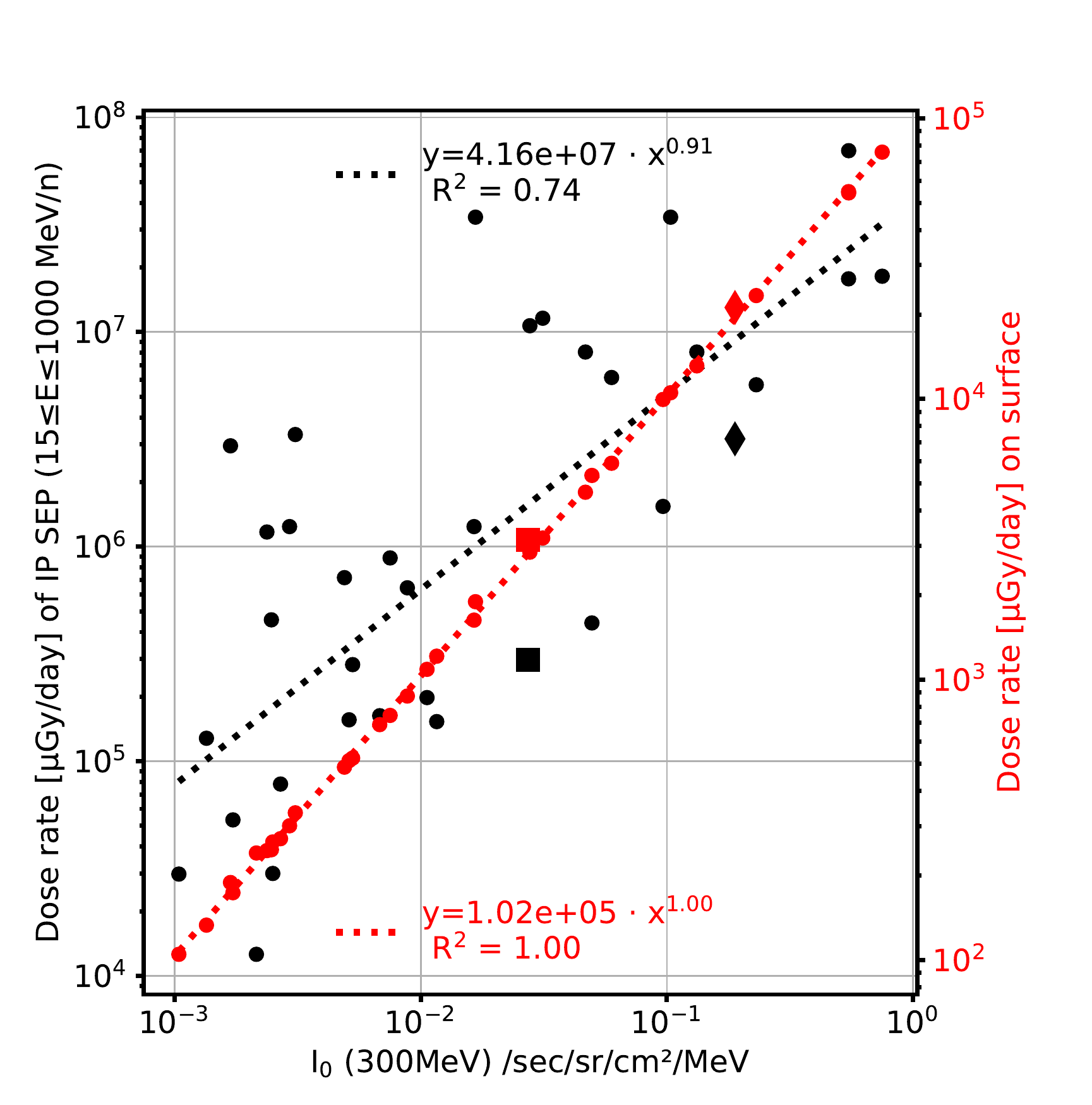}} & 
\subfloat[SEP proton integrated flux from 15 to 1000 MeV] {\includegraphics[trim=0 0 510 0, clip, scale=0.5] {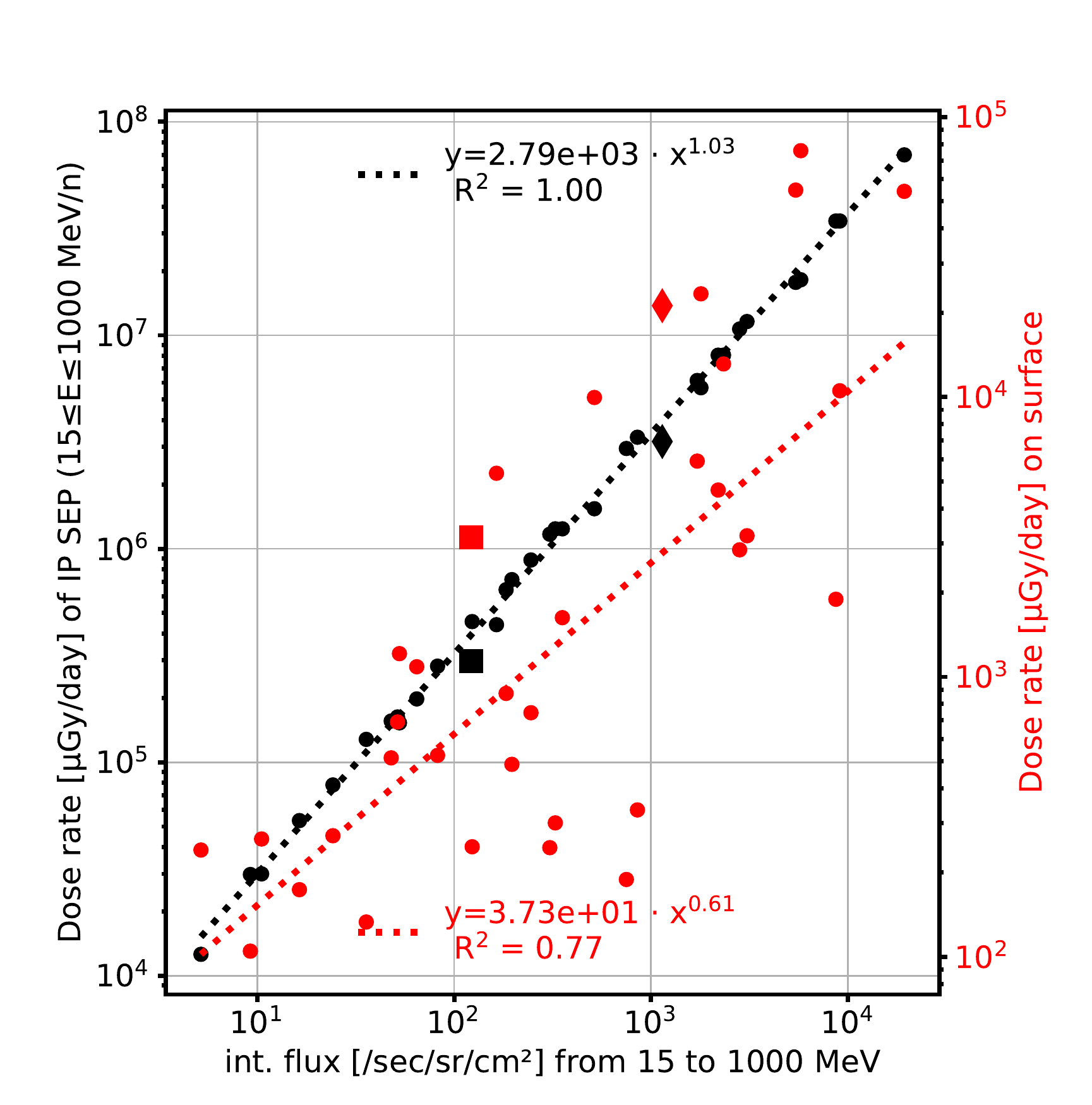}}
\end{tabular}
\caption{Case (b, primary proton from 15 to 1000 MeV) results: deep space (black, left y-axes) and Martian surface (red, right y-axes) dose rate versus SEP flux $I_{\epsilon_0}$ at ${\epsilon_0}=300 MeV$ (panel a) and the integrated flux from 15 to 1000 MeV (panel d). 
EPHIN events are marked in circular dots; Oct89 and Sep89 events are shown in squares and diamonds respectively. More explanations of the figure are shown in text of Section \ref{sec:EPHIN}. } \label{fig:doseVSI0-gamma-caseb}
\end{figure}

Fig. \ref{fig:doseVSI0-gamma-caseb} summarizes the results of case (b), where the ranges of the primary SEP protons were expanded to 15-1000 MeV {as shown in Fig. \ref{fig:EPHINspectra}}.  
The calculated dose rates in deep space (in black and left axes) and on Martian surface (in red and right axes) are plotted versus the SEP flux $I_{\epsilon_0}$ at ${\epsilon_0}=200$ MeV (panel a) and the integrated flux (panel b) of the power-law SEP spectra. 
Panel (a) shows a weaker correlation between deep space dose rate and $I_{\epsilon_0}$ (R$^2$ is 0.74) in comparison to Fig. \ref{fig:doseVSI0-gamma-casea}(c) (R$^2$ is 0.98). This is likely due to the inclusion of lower-energy protons which contribute significantly to the free-space dose with their high $dE/dx$. A smaller contribution comes from the higher-energy protons ($\geq 800$ MeV), which approach the minimum ionizing part of the $dE/dx$ curve. 
The plots and fitted parameters indicate that with an SEP power-law spectra of 15-1000 MeV, the expected proton dose rate in deep space is about 3 orders of magnitudes larger than that from 100-800 MeV protons only. 

The dose rate on the Martian surface in case (b) continues to correlate well with $I_{\epsilon_0}$ at 300 MeV and the fitted parameter (y=1.02$\times 10^5$x) are within 5\% difference compared to that in case (a) shown in Fig. \ref{fig:doseVSI0-gamma-casea}(c) (y=9.72$\times 10^4$x). 
The high similarity of surface dose rates calculated in case (a) and case (b) means that the surface radiation environment from SEPs depends mostly on primary protons in the energy range of 100-800 MeV since the lower energy protons ($\leq 100$ MeV) barely contribute to the surface dose, while the higher energy part contributes little in a power-law distribution.
The correlation between the surface dose rate and the integrated flux of the 15-1000 MeV SEP spectra in panel (b) is worse in comparison to that shown in Fig. \ref{fig:doseVSI0-gamma-casea}(d) for the same reason: a large share of the 15-1000 MeV protons makes a negligible contribution to the surface radiation environment. 
But we note that the deep space dose rate from protons in the range of 15 to 1000 MeV correlates rather well with the integrated flux of the spectra in this range.  
  
To validate the correlation derived from power-law SEPs, we compare these results with that from the historical events presented in Section \ref{sec:Storia}. 
To do so, we fitted the Oct89 and Sep89 events with power-law spectra in the energy range of 100-800 MeV, using the fitted $I_0$ and $\gamma$ shown in Fig.\ref{fig:IPspectra}. 
We have re-calculated the deep space and surface dose rates of these two events in the primary energy range of (a) 100-800 MeV and (b) 15-1000 MeV and plotted them versus the corresponding $I_{\epsilon_0}$ at 200 and 300 MeV. {We note that the calculation was performed using the event spectra within this range, not the power-law fitted spectra in order to verify the robustness of the power-law assumption.}
These results are also shown in Figs. \ref{fig:doseVSI0-gamma-casea} and \ref{fig:doseVSI0-gamma-caseb} with squares indicating the Oct89 event and diamonds standing for the Sep89 event. 
Both events are highly consistent with the events observed by EPHIN, showing the quasi-linear correlation of dose rate and $I_{\epsilon_0}$ validating the correlations derived above. 
The values of dose rates calculated from these events in cases (a) and (b) are also shown in Table \ref{table:3events_sum}. 
It is important to note that when comparing the surface dose rate induced by (a) 100-800 MeV protons and (b) 15-1000 MeV ones to that from the full SEP spectra (f), the difference is fairly small. For both Oct89 and Sep89 events, the surface dose rate from case (a) is about 93\% of that from (f); the surface dose rate from case (b) is about 96\% of that from (f). This again proves that the surface dose rate depends mostly on primary protons in the energy range of 100-800 MeV. 
 
The empirical correlation shown in Figs. \ref{fig:doseVSI0-gamma-casea} (a) and \ref{fig:doseVSI0-gamma-caseb} (a)
can be used for quick estimations of the expected dose rates both in deep space and on the surface of Mars upon the onset of a sudden solar particle event whose spectra has roughly a power-law distribution in the concerned energy range.

\section{The potential extra contribution by $^4$He ions}\label{sec:alpha}
\begin{figure}
\centering
\begin{tabular}{cc}
\subfloat[Solar $^4$He energy range: 100-800 MeV/nuc]{\includegraphics[trim=10 0 500 0, clip,scale=0.5]{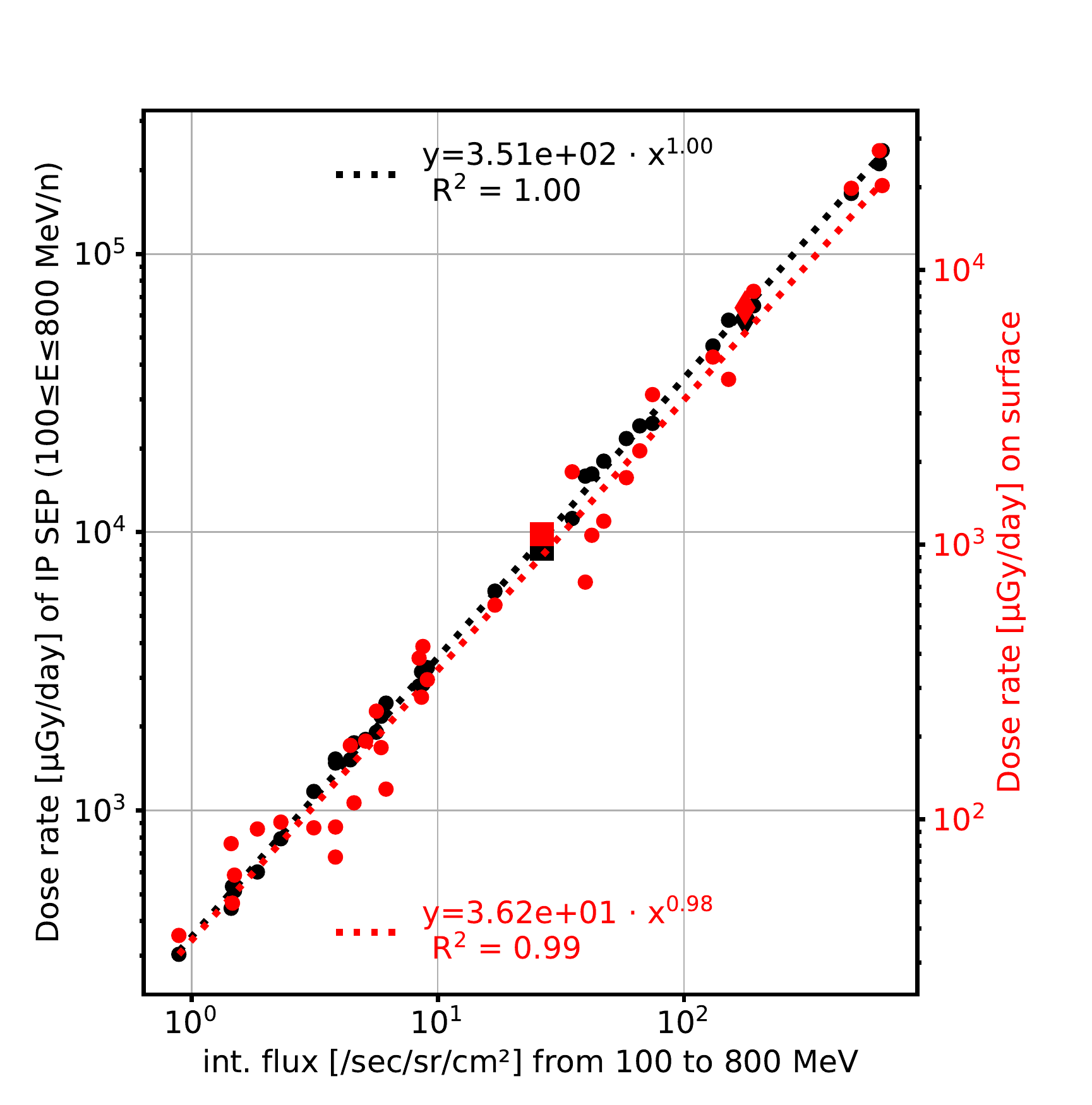}}&
\subfloat[Solar $^4$He energy range: 15-1000 MeV/nuc]{\includegraphics[trim=10 0 500 0, clip,scale=0.5]{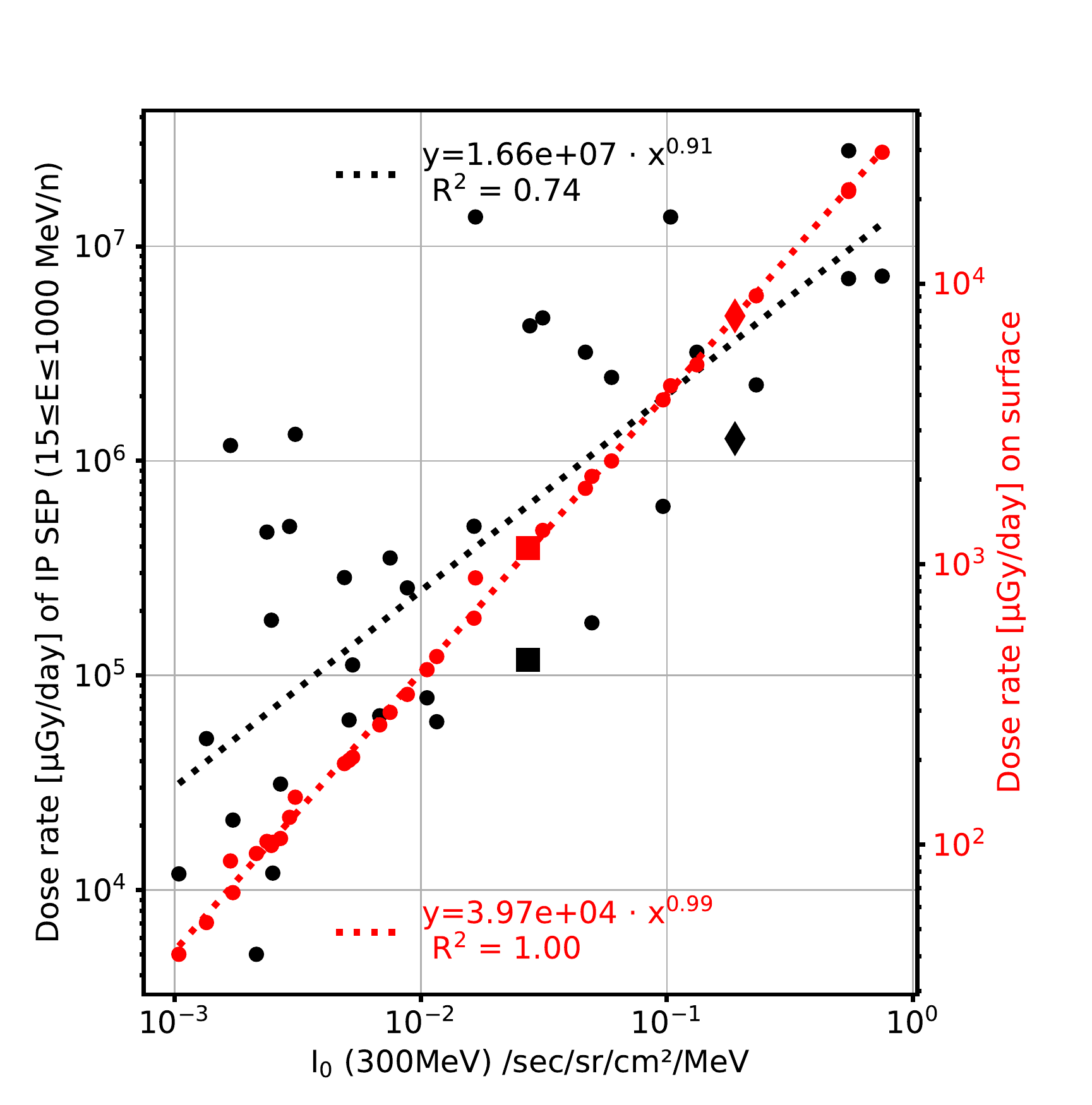}}
\end{tabular}
\caption{Deep space (black, left y-axes) and Martian surface (red, right y-axes) dose rates from solar $^4$He particles with energy range of (case a) 100-800 MeV/nuc and (case b) 15-1000 MeV/nuc. The x-axes show the integrated flux (left panel) and $I_{\epsilon_0}$ at 300 MeV (right panel) of the proton power-law spectra (used for scaling to obtain the $^4$He spectra). More explanations can be found in the text of Section \ref{sec:alpha}.
EPHIN events are marked in circular dots; Oct89 and Sep89 events are shown in squares and diamonds. 
} \label{fig:doseVSI0-gamma-alpha}
\end{figure}

Although protons are the large majority of the primary particles reaching the top of the Martian atmosphere, energetic helium ions can also propagate into deep space and a flux ratio of He/p to be about 10\% has been estimated to be a reasonable worst-case scenario based on SOHO/ERNE measurements \citep{torsti1995}. 
Based on this assumption, we have also scaled the EPHIN proton power-law fitted spectra to 1 order of magnitude smaller representing the $^4$He spectra which is then used to (1) calculate the deep space induced dose rates and (2) multiply the matrices for deriving surface spectra and dose rate from all secondaries induced by these primary He particles. This may be an unrealistic assumption since the charge-to-mass ratio of $^4$He ions makes them more difficult to accelerate than $^1$H. They will therefore tend to have softer energy spectra than protons accelerated by the same mechanism, so a simple scaling of the proton flux is likely to overestimate the contributions of $^4$He.

A summary figure similar to Figs. \ref{fig:doseVSI0-gamma-casea} and \ref{fig:doseVSI0-gamma-caseb} is shown in Fig. \ref{fig:doseVSI0-gamma-alpha}; the fitted parameters are again labeled. 
To give a direct comparison of the fitted parameters, we plotted and fitted the $^4$He-induced dose rate to the same $I_{\epsilon_0}$ of the proton spectra before scaled to $^4$He. 
Fig. \ref{fig:doseVSI0-gamma-alpha}(a) can be compared with Fig. \ref{fig:doseVSI0-gamma-casea}(d) while Fig. \ref{fig:doseVSI0-gamma-alpha}(b) can be compared with Fig. \ref{fig:doseVSI0-gamma-caseb}(a).
For unshielded deep space, the proton-induced dose rate is about 2.5 times larger than He-induced dose rate in both cases (a) and (b) as shown in by the fitted parameters labeled in black. 
This is exactly a trivial result: scaling down the proton spectrum by a factor of 10 is partially compensated by the factor of 4 higher $dE/dx$ of a $^4$He ion at the same velocity ($Z^2$ scaling) yields a factor of 2.5 in the ratio of doses. 
The Martian surface case is more complicated owing to the different transport properties of $^1$H and $^4$He.
The primary solar proton and helium induced surface dose ratio is about 2.6 for both cases, slightly larger than, but still close to, 2.5. 
This is because a fraction of the helium ions will undergo nuclear fragmentation in the atmosphere, reducing their contribution to the surface dose.
As noted above, simply scaling the proton spectrum to the $^4$He spectrum may lead to an over-estimation of the $^4$He contributions. Precise measurements of He spectra by, e.g., PAMELA \citep{picozza2007pamela} are needed for better estimations of the He induced dose rates.

\section{Discussion and Conclusion}
\begin{figure}
\centering
\includegraphics[scale=0.6]{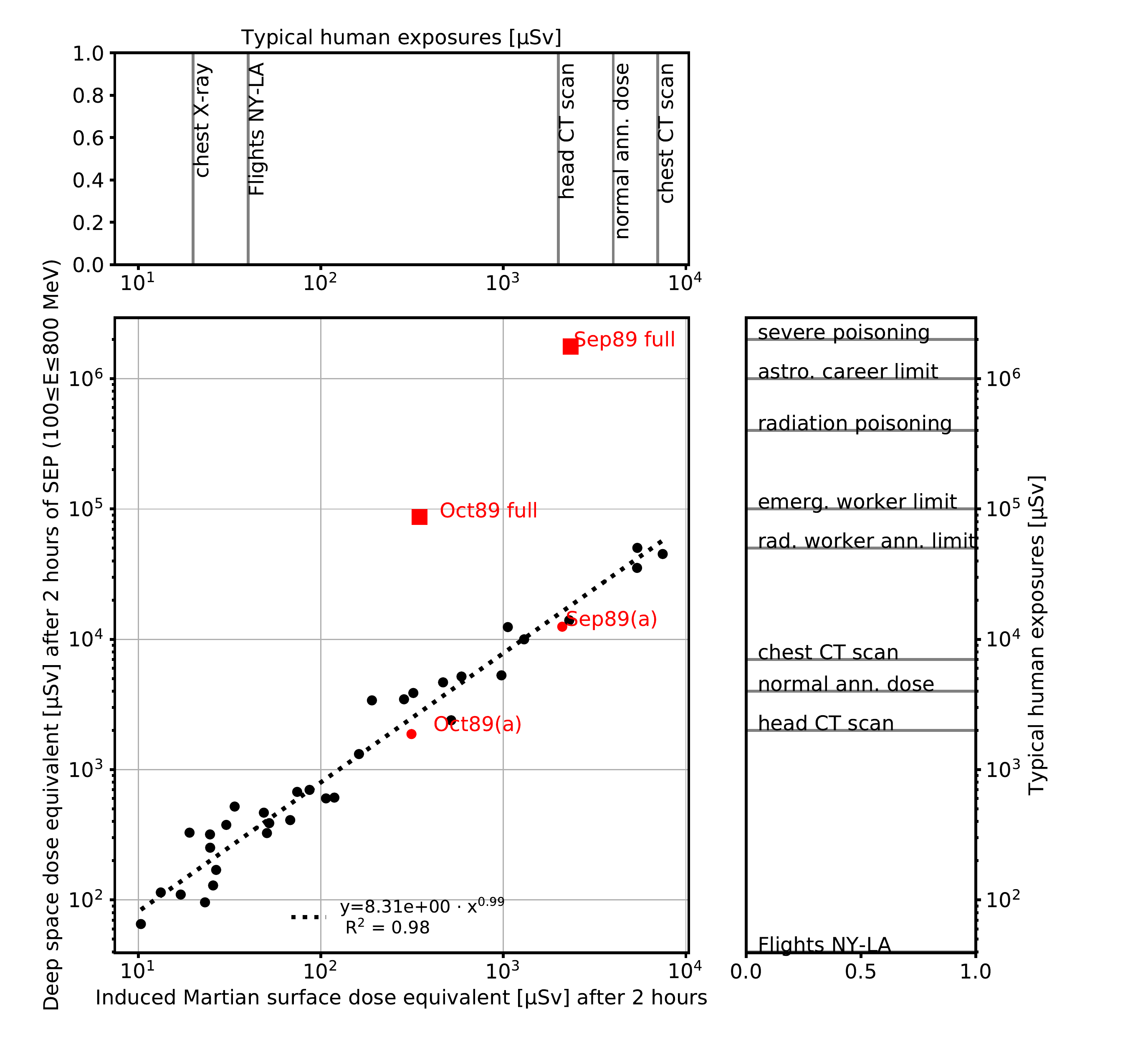}
\caption{Lower left: deep space dose equivalent versus Martian surface dose equivalent resulted from two-hour integrated power-law shaped SEPs (black dots) where primary proton energies are obtained between 100 and 800 MeV as in case (a). 
The two historical events are also plotted with circles standing for case (a) calculations and square dots for the full spectra (protons larger than 5 MeV) modeling results. 
Lower right and upper left charts mark the typical human exposures [\textmu Sv] in daily life, regulations and urgent cases taken from https://xkcd.com/radiation/. }\label{fig:dose_charts}
\end{figure}

In terms of biological effectiveness associated with radiation exposures on human beings, the dose equivalent (in units of Sv) is often more referred to for evaluating the deep space exploration risks \citep{sievert1959}. It can be computed using the linear energy transfer (LET) dependent quality factor, Q(L), from \citet{icrp60}. For LET less than 10 keV/$\mu$m in water, Q is identically 1; this value applies to the large majority of SEP protons, so the dose rates reported here are in most cases close to the corresponding dose equivalent rates. 
{As noted earlier, particles with very low energies may contribute significantly to dose and dose equivalent rates. In order to make a realistic and meaningful estimation of the biological effects, we calculated the dose equivalent rate with a low energy cutoff requiring only particles which could penetrate through 0.5 mm of tissue (e.g., protons $\ge$ 5 MeV) which is the thinnest skin of a human body (eyelids).}

We integrated such calculated dose equivalent rate for two hours for each event since the EPHIN event spectra was calculated in two-hour intervals shortly after the event onset (see Section \ref{sec:EPHIN}). Fig. \ref{fig:dose_charts} (lower-left panel) shows the dose equivalent for deep space case (y-axis, from primary protons of 100-800 MeV energy range) versus dose equivalent on surface of Mars (x-axis, all secondaries induced by primary protons of 100-800 MeV energy range).
The black dots represent the dose equivalent from all EPHIN events and the correlation coefficient between the deep space and surface dose equivalents is 0.98 and they depend on each other roughly following a simple linear relationship 
which indicates the dose equivalent rate of such events on the surface is generally 8-9 times smaller than that (from 100-800 MeV protons) in deep space. 
A similar fitting for case (b) where 15-1000 MeV primary protons were considered shows that the deep space dose equivalent rate is about 90 times larger than that on the Martian surface. 
We have omitted the contributions by $^4$He ions here, since the intensity and spectra we modeled above are speculative; in a worst-case scenario, we might expect an additional 40\% contribution from these ions.

To assess the differences between dose equivalents induced by the full spectra and the energy-limited power-law spectra, we adopted the two historical events and compared the modeled results between full spectra (a different cutoff energy at 5 MeV is applied compared to 1 MeV for the dose calculations in Section \ref{sec:Storia}) and 100-800 MeV range (case a) shown in red in Fig.\ref{fig:dose_charts}. The squares stand for the results from the full spectra (f) while the circles represent those from case (a).

For deep space case, dose equivalents from (f) and (a) differ significantly, by as much as 2 orders of magnitude.
Slightly bigger differences were recorded for dose rate ratios, as listed in Table \ref{table:3events_sum} in the row "(ad)/(fd) ratio" where, e.g., the dose rate in deep space resulted from 100-800 MeV protons is only about 0.66\% of the total dose rate (of protons larger than 1 MeV) for the Oct89 event. This ratio is even less (0.07\%) for the Sep89 event where low-energy protons are up to a couple of magnitudes more abundant as shown in Fig. \ref{fig:IPspectra}.
The table also shows the dose rates in case (b) where primary protons with energies from 15 to 1000 MeV are considered.
Such ratios shown in the row "(bd)/(fd) ratio" become 9.03\% and 1.60\% for the Oct89 and Sep89 events respectively. 
These values are larger, mainly due to the contributions of low energy protons not considered in case (a). 

For the Martian surface, the dose or dose equivalent rates do not depend significantly on the full primary spectra. 
As shown in Fig. \ref{fig:dose_charts} by the two historical events, the induced surface dose equivalents (x-axis) are very similar from the two different primary spectra (a) and (f).
The values of dose rates for Martian surface scenario are also shown in Table \ref{table:3events_sum} and the surface dose rate resulted from 100-800 MeV protons is about 93\% of the total surface dose rate for both the Oct89 and Sep89 events listed in the row "(as)/(fs) ratio".
Because of the atmospheric shielding, the contributions of 100-800 MeV protons dominate the SEP-induced environment on the surface of Mars, despite the fact that these protons contribute little to the free-space dose equivalent. 

Typical human exposures [\textmu Sv] expected in daily life and defined in regulations for special cases are marked in the lower right and the upper left panels of Fig. \ref{fig:dose_charts} as a reference for possible potential biological effects of the SEPs studied here.
For SEP events encountered in deep space without any additional shielding around, the accumulated dose equivalent for e.g., the Sep89 event after two hours would be 1.64 $\times 10^6$ \textmu Sv, a value higher than the astronaut career limit $10^6$ \textmu Sv.
However this is an over estimation since at least the space suit shielding should be present for worse-case extravehicular activities (EVA).
Since the current paper is mainly focused on consequences of the extreme events for the Martian surface case considering the Martian atmospheric shielding, we will not go into details discussing about the deep space scenarios. 
Interested readers are pointed to previous studies by e.g., \citep{wilson2006spacesuit} who have carried out more detailed investigations of the dose and dose equivalent responses as a function of primary proton energies considering scenarios of EVA and within spacecraft shielding conditions. 

On the surface of Mars, the dose equivalents induced by all studied events (which are significant events detected by SOHO/EPHIN over two decades) for the duration of two hours are below 10$^4$ \textmu Sv, a value well below the limit of radiation worker annual limits. 
Exposer to the Sep89 event for 2 hours would have an effect of approximately a head CT scan. 
These values are calculated for the surface of Mars at -4.4 km elevation (Gale Crater), where the atmospheric column depth averages about 22 g cm$^{-2}$. 
A habitat covered by $\sim$ 10 cm of Martian soil would provide important additional shielding against energetic particles reaching the surface.
Alternatively, spacesuits would already provide a slight protection against low-energy particles.
Detailed studies would involve further modeling of the shielding response function (by a similar matrix-set) of the spacesuit and shelter materials and will be carried out in our future work. 
Nevertheless, the current study has provided some benchmark and convenient formulas for estimating the Martian surface radiation environment induced by power-law shaped SEPs. 
The results highlight the need for future astronauts on the surface of Mars to receive space weather forecasts, and to carry alarming dosimeters \citep{NASA2014} so that they can seek an emergency shelter should a hard-spectrum event reach Mars.   
For better space weather forecasts and predicting the arrival of such hazardous events, we emphasize the importance of a space weather monitoring package including a particle detector to be embarked in all planetary and astronomical missions as a basic payload requirement.

\clearpage
\acknowledgments
The work is supported by DLR and DLR's Space Administration grant numbers 50QM0501, 50QM1201 and 50QM1701 to the Christian Albrechts University, Kiel. 

\bibliographystyle{apj}

\end{document}